\documentclass[12pt]{article}
\usepackage{graphics}
\usepackage{epsf}
\begin{document}


\newcommand{\be}{\begin{equation}}
\newcommand{\beq}{\begin{equation}}
\newcommand{\eeq}{\end{equation}}
\newcommand{\ee}{\end{equation}}

\newcommand{\beqn}{\begin{eqnarray}}
\newcommand{\eeqn}{\end{eqnarray}}
\newcommand{\bea}{\begin{eqnarray}}
\newcommand{\ena}{\end{eqnarray}}
\newcommand{\ra}{\rightarrow}
\newcommand{\susy}{{{\cal SUSY}$\;$}}
\newcommand{\su}{$ SU(2) \times U(1)\,$}

\newcommand{\gag}{$\gamma \gamma$ }
\newcommand{\gam}{\gamma \gamma }
\def\W{{\mbox{\boldmath $W$}}}
\def\B{{\mbox{\boldmath $B$}}}
\def\V{{\mbox{\boldmath $V$}}}
\newcommand{\np}{Nucl.\,Phys.\,}
\newcommand{\pl}{Phys.\,Lett.\,}
\newcommand{\pr}{Phys.\,Rev.\,}
\newcommand{\prl}{Phys.\,Rev.\,Lett.\,}
\newcommand{\prep}{Phys.\,Rep.\,}
\newcommand{\zp}{Z.\,Phys.\,}
\newcommand{\sovjnp}{{\em Sov.\ J.\ Nucl.\ Phys.\ }}
\newcommand{\nuclinst}{{\em Nucl.\ Instrum.\ Meth.\ }}
\newcommand{\annp}{{\em Ann.\ Phys.\ }}
\newcommand{\intjmp}{{\em Int.\ J.\ of Mod.\  Phys.\ }}

\newcommand{\eps}{\epsilon}
\newcommand{\mw}{M_{W}}
\newcommand{\mww}{M_{W}^{2}}
\newcommand{\mwmw}{M_{W}^{2}}
\newcommand{\mhmh}{M_{H}^2}
\newcommand{\mz}{M_{Z}}
\newcommand{\mzz}{M_{Z}^{2}}

\newcommand{\lra}{\leftrightarrow}
\newcommand{\tr}{{\rm Tr}}
\def\ls1{{\not l}_1}
\newcommand{\cms}{centre-of-mass\hspace*{.1cm}}

\newcommand{\dkg}{\Delta \kappa_{\gamma}}
\newcommand{\dkz}{\Delta \kappa_{Z}}
\newcommand{\dz}{\delta_{Z}}
\newcommand{\dgz}{\Delta g^{1}_{Z}}
\newcommand{\dgzt}{$\Delta g^{1}_{Z}\;$}
\newcommand{\la}{\lambda}
\newcommand{\lag}{\lambda_{\gamma}}
\newcommand{\lambdae}{\lambda_{e}}
\newcommand{\laz}{\lambda_{Z}}
\newcommand{\lnl}{L_{9L}}
\newcommand{\lnr}{L_{9R}}
\newcommand{\lt}{L_{10}}
\newcommand{\lu}{L_{1}}
\newcommand{\ld}{L_{2}}
\newcommand{\cw}{\cos\theta_W}
\newcommand{\sw}{\sin\theta_W}
\newcommand{\tw}{\tan\theta_W}
\def\cww{\cos^2\theta_W}
\def\sww{\sin^2\theta_W}
\def\tww{\tan^2\theta_W}

\newcommand{\epm}{$e^{+} e^{-}\;$}
\newcommand{\epemt}{$e^{+} e^{-}\;$}
\newcommand{\epem}{e^{+} e^{-}\;}
\newcommand{\ememt}{$e^{-} e^{-}\;$}
\newcommand{\emem}{e^{-} e^{-}\;}
\newcommand{\eeww}{e^{+} e^{-} \ra W^+ W^- \;}
\newcommand{\eewwt}{$e^{+} e^{-} \ra W^+ W^- \;$}
\newcommand{\epemww}{e^{+} e^{-} \ra W^+ W^- }
\newcommand{\epemwwt}{$e^{+} e^{-} \ra W^+ W^- \;$}
\newcommand{\eennhht}{$e^{+} e^{-} \ra \nu_e \bar \nu_e HH\;$}
\newcommand{\eennhh}{e^{+} e^{-} \ra \nu_e \bar \nu_e HH\;}
\newcommand{\ppwg}{p p \ra W \gamma}
\newcommand{\wwhh}{W^+ W^- \ra HH\;}
\newcommand{\wwhht}{$W^+ W^- \ra HH\;$}
\newcommand{\ppwz}{pp \ra W Z}
\newcommand{\ppwgt}{$p p \ra W \gamma \;$}
\newcommand{\ppwzt}{$pp \ra W Z \;$}
\newcommand{\gamgamt}{$\gamma \gamma \;$}
\newcommand{\gamgam}{\gamma \gamma \;}
\newcommand{\egamt}{$e \gamma \;$}
\newcommand{\egam}{e \gamma \;}
\newcommand{\gamgamwwt}{$\gamma \gamma \ra W^+ W^- \;$}
\newcommand{\gamgamwwht}{$\gamma \gamma \ra W^+ W^- H \;$}
\newcommand{\gamgamwwh}{\gamma \gamma \ra W^+ W^- H \;}
\newcommand{\gamgamwwhht}{$\gamma \gamma \ra W^+ W^- H H\;$}
\newcommand{\gamgamwwhh}{\gamma \gamma \ra W^+ W^- H H\;}
\newcommand{\ggww}{\gamma \gamma \ra W^+ W^-}
\newcommand{\ggwwt}{$\gamma \gamma \ra W^+ W^- \;$}
\newcommand{\ggwwht}{$\gamma \gamma \ra W^+ W^- H \;$}
\newcommand{\ggwwh}{\gamma \gamma \ra W^+ W^- H \;}
\newcommand{\ggwwhht}{$\gamma \gamma \ra W^+ W^- H H\;$}
\newcommand{\ggwwhh}{\gamma \gamma \ra W^+ W^- H H\;}
\newcommand{\ggwwz}{\gamma \gamma \ra W^+ W^- Z\;}
\newcommand{\ggwwzt}{$\gamma \gamma \ra W^+ W^- Z\;$}
\def\smx{{\cal{S}}{\cal{M}}}

\newcommand{\ptu}{p_{1\bot}}
\newcommand{\vecptu}{\vec{p}_{1\bot}}
\newcommand{\ptd}{p_{2\bot}}
\newcommand{\vecptd}{\vec{p}_{2\bot}}
\newcommand{\ie}{{\em i.e.}}
\newcommand{\cm}{{{\cal M}}}
\newcommand{\cl}{{{\cal L}}}
\newcommand{\cd}{{{\cal D}}}
\newcommand{\cv}{{{\cal V}}}
\def\slashc{c\kern -.400em {/}}
\def\slashp{p\kern -.400em {/}}
\def\slashL{L\kern -.450em {/}}
\def\slashcl{\cl\kern -.600em {/}}
\def\Ww{{\mbox{\boldmath $W$}}}
\def\B{{\mbox{\boldmath $B$}}}
\def\noi{\noindent}
\def\nn{\noindent}
\def\sm{${\cal{S}} {\cal{M}}\;$}
\def\smn{${\cal{S}} {\cal{M}}$}
\def\nph{${\cal{N}} {\cal{P}}\;$}
\def\sb{$ {\cal{S}}  {\cal{B}}\;$}
\def\ssb{${\cal{S}} {\cal{S}}  {\cal{B}}\;$}
\def\ssbe{{\cal{S}} {\cal{S}}  {\cal{B}}}
\def\cviol{${\cal{C}}\;$}
\def\pviol{${\cal{P}}\;$}
\def\cpviol{${\cal{C}} {\cal{P}}\;$}

\newcommand{\lgg}{\lambda_1\lambda_2}
\newcommand{\lww}{\lambda_3\lambda_4}
\newcommand{\ppin}{ P^+_{12}}
\newcommand{\pmin}{ P^-_{12}}
\newcommand{\ppout}{ P^+_{34}}
\newcommand{\pmout}{ P^-_{34}}
\newcommand{\sinsq}{\sin^2\theta}
\newcommand{\cossq}{\cos^2\theta}
\newcommand{\yt}{y_\theta}
\newcommand{\hppll}{++;00}
\newcommand{\hpmll}{+-;00}
\newcommand{\hpplt}{++;\lambda_30}
\newcommand{\hpmlt}{+-;\lambda_30}
\newcommand{\hpptt}{++;\lambda_3\lambda_4}
\newcommand{\hpmtt}{+-;\lambda_3\lambda_4}
\newcommand{\dk}{\Delta\kappa}
\newcommand{\klam}{\Delta\kappa \lambda_\gamma }
\newcommand{\kac}{\Delta\kappa^2 }
\newcommand{\lac}{\lambda_\gamma^2 }
\def\gamgamtzz{$\gamma \gamma \ra ZZ \;$}
\def\gamgamtww{$\gamma \gamma \ra W^+ W^-\;$}
\def\gamgamtwwe{\gamma \gamma \ra W^+ W^-}





\begin{titlepage}

\def\baselinestretch{1.2}

\topmargin     -0.25in

\vspace*{\fill}
\begin{center}
{\large {\bf Bosonic Quartic Couplings at LEP2 }} \vspace*{0.5cm}

\begin{tabular}[t]{c}

{\bf G.~B\'elanger$^{1}$, F.~Boudjema$^{1}$, Y.~Kurihara$^{2}$,
D.~Perret-Gallix$^{3}$, A.~Semenov$^{4}$ }
 \\
\\
\\

{\it 1. Laboratoire de Physique Th\'eorique} {\large
LAPTH}\footnote{URA 14-36 du CNRS, associ\'ee  \`a l'Universit\'e
de Savoie.}\\
 {\it Chemin de Bellevue, B.P. 110, F-74941 Annecy-le-Vieux,
Cedex, France.}\\
\\


{\it 2. High Energy Accelerator Research Organisation, {\large
KEK},  }\\ {\it Tsukuba, Ibaraki 305-801, Japan.}\\
\\


{\it 3. LAPP} \\ {\it Chemin de Bellevue, B.P. 110, F-74941
Annecy-le-Vieux, Cedex, France. }\\
\\

{\it 4. Joint Institute for Nuclear Research, {\large JINR}, } \\
{\it 141980 Dubna, Russia. }

\end{tabular}
\end{center}

\centerline{ {\bf Abstract} }

\baselineskip=14pt
\noindent
 {\small We list the set of \cviol and \pviol conserving anomalous quartic vector bosons
self-couplings which can be tested at LEP2 through triple vector
boson production. We show how this set can be embedded in
manifestly $SU(2) \times U(1)$ gauge invariant operators
exhibiting an $SU(2)_c$ global symmetry.
 We derive bounds on these various couplings and show the most relevant
 distributions that can
 enhance their contribution. We also find that an $e^+ e^-$ collider
 running at $500$~GeV  can improve the LEP2 limits by as much as three-orders of magnitude.
}
\vspace*{\fill}

\vspace*{0.1cm}
\rightline{KEK-CP-087}
\rightline{LAPTH-744/99}
\end{titlepage}
\baselineskip=18pt

\setcounter{section}{1}

\setcounter{subsection}{0}
\setcounter{equation}{0}
\def\thesubsection {\thesection.\arabic{subsection}}
\def\theequation{\thesection.\arabic{equation}}

\setcounter{equation}{0}
\def\thequation{\thesection.\arabic{equation}}

\setcounter{section}{0} \setcounter{subsection}{0}

\def\eewwgt{$e^+ e^-\ra W^+ W^- \gamma \;$}
\def\eewwg{$e^+ e^-\ra W^+ W^-$}
\section{Photonic Quartic Couplings}

LEP has now crossed the threshold for Z pair production and
therefore experiments can now study triple boson production like
$\epem \ra W^+ W^- \gamma, Z Z\gamma$, beside  $\epem \ra \gamma
\gamma \gamma, \gamma \gamma Z$ which may be studied at lower
energies. These processes have the potential to study new quartic
photonic couplings, photonic in the sense that at least one of the
vector bosons is a photon. One should refer to these quartic
couplings as {\em genuine} quartic couplings
\cite{nousggvv,Nous3gamma,nousee3v} contrary to quartic couplings
that may emerge from an operator that induces for instance both a
tri-linear $WW\gamma$ coupling as well as a possible $WW\gamma
\gamma$, as required by gauge invariance. The latter (non-genuine)
couplings can therefore be investigated much more efficiently
through their tri-linear counterpart in, for instance, $\epem \ra
W^+ W^-$. An example of such a coupling is the much studied
operator described by $\lambda_\gamma$ in the by now classic
classification \cite{HPZH} , $\lambda_\gamma$ is sometimes
referred to as the anomalous quadrupole moment of the $W$. From
this perspective genuine quartic couplings can only be studied in
triple vector boson production or through boson-boson fusion, the
latter becoming a more  efficient means at TeV energies
\cite{nousee3v}.

Let us note that quartic {\em neutral} couplings, $4-\gamma\;, \;
Z-3\gamma$, contributing to $\epem \ra 3\gamma$ have already been
studied in \cite{Nous3gamma} and could be explored for energies
below those presently available. Since these quartic couplings
involve at least three photons, electromagnetic gauge invariance
alone allows these couplings only if they emerge from
dimension-eight (or higher) operators. On the other hand,
anomalous couplings such as $WW\gamma\gamma$ or $WWZ\gamma$ that
contribute to \eewwgt may be associated to dim-6 operators and are
hence a more likely signal of a possible residual effects of New
Physics. As a matter of fact $WW\gamma\gamma$ and $WWZ\gamma$ are
present in the \sm at tree-level and as a consequence these types
of couplings are more important to study. These quartic photonic
couplings were first introduced in \cite{nousggvv} in view of
studying their effects on \ggwwt in the laser mode of the Linear
Collider.  They were  derived by only appealing to electromagnetic
gauge invariance and $SU(2)_c$ custodial symmetry. The
phenomenology of these couplings has since then been studied in
the next linear collider both in the \epemt
\cite{nousee3v,Stirlingquartic1}, $\gamma\gamma$ \cite{nousggvv}
and $e \gamma$ \cite{eboliquartic} modes. Very recently these
couplings have been re-investigated for LEP2 energies
\cite{Stirlingquartic}. Unfortunately, as we will show, when
studying the effect of genuine quartic couplings in \eewwgt and
$\epem \ra Z\gamma\gamma$, one needs to consider a larger set of
structures than the two that have been written down for \ggwwt.
The aim of this paper is to generalize the study we performed in
\cite{nousggvv,nousee3v} and to review and clarify some of the
issues related to the photonic quartic couplings.

The plan of the paper is as follows. In the next section, we start
by first listing the leading quartic operators that contribute to
$\epem \ra W^+W^-\gamma$ (and $Z\gamma\gamma$). In writing down
this list we will only appeal to explicit $U(1)_{\rm em}$ gauge
invariance as well as \cviol and \pviol conservation. In a sense
these structures constitute the quartic counterpart to the
tri-linear classification in \cite{HPZH}. In passing we will point
out that a third ``photonic" quartic coupling that has been
entertained \cite{eboliquartic,Stirlingquartic} in the literature
does in fact violate \cpviol.  We will then show how the different
structures can be embedded within $SU(2) \times U(1)$ operators
which we require also to exhibit the $SU(2)_c$ global custodial
symmetry which leads to $\rho=1$ in the limit of vanishing
hypercharge coupling. This can be done either in the usual
approach by exploiting the covariant derivative on the
Goldstone-Higgs field (for notations and conventions refer to
\cite{Morioka,ChopinHiggs}) or in the non-linear chiral  approach
of symmetry breaking (see \cite{Morioka,nousggwwNP}). The
explicitly  $SU(2)\times U(1)$ approach together with the $SU(2)$
global symmetry will allow to relate some $WW\gamma \gamma$ and
$ZZ\gamma \gamma$ structures,  for example. In section~3 we turn
to the analysis of these quartic couplings in \eewwgt and $\epem
\ra Z\gamma \gamma$. We will derive the limits  one may hope to
extract and show the distributions which are most sensitive to
these couplings. The case of the \cpviol violating operator is
relegated to an Appendix.

\section{Structures which contribute to \eewwgt, $Z\gamma \gamma$ and $ZZ\gamma$}

For LEP2, the processes of interest, and the lowest-dimension
anomalous quartic couplings they are sensitive to, are
\begin{itemize}
\item $ \epem \ra W^+ W^- \gamma \;\;\;\;\; \longrightarrow {\rm quartic:}\;\; WW\gamma \gamma, WWZ \gamma$
\item $ \epem \ra Z \gamma \gamma \;\;\;\;\; \longrightarrow {\rm quartic}\;\; Z Z\gamma \gamma$
\item $\epem \ra Z Z \gamma \;\;\;\;\; \longrightarrow {\rm quartic} \;\; ZZZ\gamma, ZZ \gamma \gamma$
\end{itemize}

Due to phase space the latter process is marginal at LEP2. Note
that for $Z\gamma \gamma$ production only one coupling is checked,
$ZZ\gamma \gamma$, if one restricts oneself to the lowest dimension
operators, otherwise a $Z 3\gamma$ which is of highest dimension
may also contribute. Already at LEP2, one may also exploit $\epem \ra \nu
\bar{\nu} \gamma \gamma$ as a testing ground for the quartic
coupling $WW\gamma \gamma$, as suggested in
\cite{Stirlingquartic,Opalquartic2}.

We start by listing all those genuine quartic bosonic operators
that contribute to the latter processes and which are of lowest
possible dimension, as it turns out, dim-6. We first only require
electromagnetic gauge invariance together with \cviol and \pviol
symmetry. At this stage the $WW\gamma\gamma$, $WWZ\gamma$ or
$ZZ\gamma \gamma$ couplings, for example, are not related. Each
photon requires the use of the electromagnetic tensor $F_{\mu
\nu}=\partial_\mu A_\nu -\partial_\nu A_\mu$.
As we have shown  elsewhere \cite{nousggvv}, there can be only two
basic Lorentz structures for the lowest dimension $WW\gamma
\gamma$ operators. These map into the parameters $a_0$ and $a_c$
first introduced in \cite{nousggvv,nousee3v}. Hence the two
$WW\gamma \gamma$ Lorentz structures are:

\beqn
\label{coup_wwgg}
{{\cal W}}_0^\gamma&=&-\frac{e^2 g^2}{2 \Lambda^2}\; F_{\mu \nu}
F^{\mu \nu} W^{+\alpha} W^-_\alpha \nonumber
\\ {{\cal W}}_c^\gamma&=&-\frac{e^2 g^2}{4 \Lambda^2}\;
F_{\mu \nu} F^{\mu \alpha} \left( W^{+\nu}
W^-_\alpha \;+\; W^{-\nu} W^+_\alpha \right)
\eeqn

where $e$ is the electromagnetic coupling,
$g=e/\sin\theta_W=e/s_W$ and $\Lambda$ a mass scale characterizing
the New Physics.

For $W^+W^-Z\gamma$, it is also easy to see that one can have a
maximum of $5$ independent structures. With $g_Z=e/s_W c_W$ and
$V_{\mu \nu}=\partial_\mu V_\nu -\partial_\nu V_\mu$ where $V=W^\pm,Z$,
we have
\beqn
\label{coup_wwzg}
{{\cal W}}_0^Z&=&-\frac{e^2 g^2}{\Lambda^2} F_{\mu \nu} Z^{\mu
\nu} W^{+\alpha} W^-_\alpha \nonumber
\\ {{\cal W}}_c^Z&=&-\frac{e^2 g^2}{2 \Lambda^2} F_{\mu \nu} Z^{\mu \alpha} \left( W^{+\nu}
W^-_\alpha \;+\; W^{-\nu} W^+_\alpha \right) \nonumber
\\{{\cal W}}_1^Z&=&-\frac{e g_Z g^2}{2 \Lambda^2}\; F^{\mu \nu} \left( W^+_{\mu \nu} W^-_\alpha Z^\alpha \;+\; W^-_{\mu \nu} W^+_\alpha Z^\alpha \right)
\nonumber
\\{{\cal W}}_2^Z&=&-\frac{e g_Z g^2}{2 \Lambda^2}\; F^{\mu \nu}\left ( W^+_{\mu \alpha} W^{-\alpha}
Z_\nu \;+\; W^-_{\mu \alpha} W^{+\alpha} Z_\nu\right) \nonumber
\\{{\cal W}}_3^Z&=&-\frac{e g_Z g^2}{2 \Lambda^2}\; F^{\mu \nu}\left ( W^+_{\mu \alpha} W^-_\nu
Z^\alpha \;+\; W^-_{\mu \alpha} W^+_\nu Z^\alpha \right) \nonumber
\nonumber
\\
\eeqn

\def\wng{{{\cal W}}_0^\gamma}
\def\wnz{{{\cal W}}_0^Z}
\def\wcg{{{\cal W}}_c^\gamma}
\def\wcz{{{\cal W}}_c^Z}
\def\wuz{{{\cal W}}_1^Z}
\def\wdz{{{\cal W}}_2^Z}
\def\wtz{{{\cal W}}_3^Z}
\def\zng{{{\cal Z}}_0^\gamma}
\def\zcg{{{\cal Z}}_c^\gamma}
\def\znz{{{\cal Z}}_0^Z}
\def\zcz{{{\cal Z}}_c^Z}

Note that instead of the use of the field tensor, $V_{\mu \nu}$,
for one of the massive vector boson in Eq.~\ref{coup_wwzg}, we
could have used instead a simple derivative, $\partial_\mu V_\nu$.
However it is easy to show that using the derivative only maps
into one/or a combination of the above $7$ operators, if one
requires the photon from Eq.~\ref{coup_wwzg} to be on-shell like
in the process of interest, \eewwgt.  Therefore, all in all, there
are 7 \cviol and \pviol conserving Lorentz structures  which at
leading order contribute to \eewwgt. Note that at high enough
energy one may differentiate, in \eewwgt, between the quartic
couplings of type  ${{\cal W}}_{0,c}$ and those of the type
${{\cal W}}_{1,2,3}^Z$  if one is able to reconstruct the final
polarisation of the $W$'s. Indeed both $W$'s in the former are
preferentially longitudinal whereas in the latter, one is
transverse and the other longitudinal, this is because in the
latter the operators involve at least a field strength to describe
a $W$.

 It is straightforward to ``convert"
the above operators to genuine quartic couplings for $ZZ\gamma
\gamma$ and $ZZZ\gamma$ which contribute to $\epem \ra Z\gamma
\gamma$ and $\epem \ra ZZ\gamma$. One counts two independent
operators for $ZZ\gamma \gamma$
\beqn
\label{coup_zzgg}
{{\cal Z}}_0^\gamma&=&-\frac{e^2 g_Z^2}{4 \Lambda^2}\; F_{\mu \nu}
F^{\mu \nu} Z^{\alpha} Z_\alpha \nonumber
\\ {{\cal Z}}_c^\gamma&=&-\frac{e^2 g_Z^2}{4 \Lambda^2}\; F_{\mu \nu} F^{\mu \alpha} Z^{\nu}
Z_\alpha
\eeqn
and two for $ZZZ\gamma$
\beqn
\label{coup_zzzg}
{{\cal Z}}_0^Z&=&-\frac{e^2 g_Z^2}{2\Lambda^2}\; F_{\mu \nu} Z^{\mu \nu} Z^{\alpha}
Z_\alpha \nonumber
\\ {{\cal Z}}_c^Z&=&-\frac{e^2 g_Z^2}{2 \Lambda^2}\; F_{\mu \nu} Z^{\mu \alpha} Z^{\nu}
Z_\alpha
\eeqn

\subsection{Feynman Rules}
The Feynman rules for the above operators are easy to derive. It is worth noticing that
all the above operators can be expressed in terms of very few Lorentz structures. We
define

\beqn
{{\cal P}}_0(A(k_1, \mu); N(k_2, \nu); V_\alpha \;; V_\beta)=
\frac{i e^2 g^2}{\Lambda^2}\;\; 2\; g_{\alpha \beta} (g_{\mu \nu}
k_1.k_2 - k_{1 \nu} k_{2 \mu})
\eeqn
$k_i$ stand for the momentum of the particle and $\{\mu \; \nu \;
\alpha\; \beta\}$ are the Lorentz indices.

\beqn
{{\cal P}}_c(A(k_1, \mu); N(k_2, \nu); V_\alpha ; V_\beta)= \frac{i e^2 g^2 }{2
\Lambda^2}\;\; \nonumber \\
\left( (g_{\mu \alpha} g_{\nu \beta}+ g_{\nu \alpha} g_{\mu \beta}) k_1.k_2 \;+\; g_{\mu
\nu} (k_{2 \beta} k_{1 \alpha}+k_{1 \beta} k_{2 \alpha}) \right.\nonumber
\\
- \left. k_{2 \mu} k_{1 \alpha} g_{\nu \beta}\;-\; k_{2 \beta} k_{1 \nu} g_{\mu
\alpha}\;-\; k_{2 \alpha} k_{1 \nu} g_{\mu \beta}\;-\; k_{2 \mu} k_{1 \beta} g_{\nu
\alpha} \right)
\eeqn

For the $WWZ\gamma$ with no $WW\gamma \gamma$ equivalent, the last
three structures in Eq.~\ref{coup_wwzg}, three more structures are
needed

\beqn
{{\cal P}}_1(A(k_1, \mu); Z(k_2, \nu); W^+(k_+,\alpha) ;
W^-(k_-,\beta) )= \frac{i e g_Z g^2 }{ \Lambda^2}\;\;\nonumber \\
\left( (k_1.k_+ g_{\mu \alpha} -k_{+ \mu} k_{1 \alpha})g_{\nu
\beta}\;+ \;(k_1.k_- g_{\mu \beta} -k_{-\mu} k_{1 \beta})g_{\nu
\alpha} \right)
\eeqn

\beqn
{{\cal P}}_2(A(k_1, \mu); Z(k_2, \nu); W^+(k_+,\alpha) ;
W^-(k_-,\beta) )= \frac{i e g_Z g^2 }{ 2 \Lambda^2}\;\;\nonumber
\\ \left( (k_1.k_+ +k_1.k_-) g_{\mu \nu} g_{\alpha
\beta}\;-\;(k_{1 \alpha} k_{+ \beta}+ k_{1 \beta} k_{-
\alpha})g_{\mu \nu} \right. \nonumber
\\-\left. ( k_{+ \mu}+k_{- \mu})k_{1 \nu} g_{\alpha \beta}\;+\;
(k_{+ \beta}g_{\mu \alpha} +k_{- \alpha}g_{\mu \beta})k_{1 \nu}\right)
\eeqn

\beqn
{{\cal P}}_3(A(k_1, \mu); Z(k_2, \nu); W^+(k_+,\alpha) ;
W^-(k_-,\beta) )= \frac{i e g_Z g^2 }{ 2 \Lambda^2}\;\; \nonumber
\\ \left( k_1.k_+ g_{\mu \beta} g_{\nu \alpha} \;+\;k_1.k_- g_{\mu
\alpha} g_{\nu \beta}\;+\;(k_{+\nu}-k_{-\nu})k_{1\beta}g_{\mu
\alpha}\right. \nonumber
\\-\left.(k_{+\nu}-k_{-\nu})k_{1\alpha}g_{\mu \beta}\;-\;k_{+\mu}k_{1\beta}g_{\nu
\alpha}\;-\;k_{-\mu}k_{1\alpha}g_{\nu \beta} \right)
\eeqn

For the couplings of four neutral bosons, it is useful to introduce
\beqn
{{\cal P}}_{0,c}^Z={{\cal P}}_{0,c}(g\rightarrow g_Z)
\eeqn

then taking all particles to be incoming, the Feynman rules are
\beqn
{{\cal W}}_{0,c}^\gamma &\rightarrow& {{\cal P}}_{0,c}(A(k_1, \mu); A(k_2, \nu);
W^+(\alpha); W^-(\beta) ) \nonumber
\\ {{\cal W}}_{0,c}^Z &\rightarrow& {{\cal P}}_{0,c}(A(k_1, \mu); Z(k_2, \nu);
W^+(\alpha); W^-(\beta) ) \nonumber
\\ {{\cal W}}_{1,2,3}^Z &\rightarrow& {{\cal P}}_{1,2,3}(A(k_1, \mu); Z(k_2, \nu);
W^+(k_+\alpha); W^-(k_-\beta) ) \nonumber
\\ {{\cal Z}}_{0,c}^\gamma &\rightarrow& {{\cal
P}}_{0,c}^Z(A(k_1, \mu); A(k_2, \nu); Z(\alpha); Z(\beta) ) \nonumber
\\
{{\cal Z}}_{0,c}^Z &\rightarrow& {{\cal P}}_{0,c}^Z(A(k, \mu); Z(k_1, \nu); Z(k_2\rho);
Z(k_3 \lambda) ) \;+\; \left((k_1, \nu)\leftrightarrow (k_2,\rho)\right) \;+\;\left((k_1,
\nu)\leftrightarrow (k_3,\lambda) \right) \nonumber \\
&&
\eeqn

\subsection{Embedding in gauge invariant $SU(2)_c$ symmetric operators}
All the above operators can be embedded in manifestly $SU(2)
\times U(1)$ gauge invariant and $SU(2)_c$ symmetric operators,
which are then \cviol and \pviol conserving. The construction has
been explained at some length in \cite{Morioka}. For these kind of
quartic operators it is more appropriate to use the chiral
Lagrangian approach, which assumes no Higgs. The leading order
operators in the energy expansion reproduce the ``Higgsless"
standard model. Introducing our notations, as concerns the purely
bosonic sector, the SU(2) kinetic term that gives the standard
tree-level gauge self-couplings is
\beqn
\cl_{{\rm Gauge}}=- \frac{1}{2} \left[ \tr(\Ww_{\mu \nu} \Ww^{\mu
\nu}) + \tr(\B_{\mu \nu} \B^{\mu \nu}) \right]
\eeqn

where the $SU(2)$ gauge fields  are $\Ww_\mu=W^{i}_{\mu}\tau^i$,
while the hypercharge field is denoted by $\B_\mu=\tau_3 B_\mu$.
The normalisation for the Pauli matrices is $\tr(\tau^i
\tau^j)=2\delta^{i j}$. We define the field strength as, $\Ww_{\mu
\nu}$
\beqn
\Ww_{\mu \nu}&=&\frac{1}{2} \left(
\partial_\mu \Ww_{\nu}- \partial_\nu \Ww_{\mu} +\frac{i}{2}
g[\Ww_\mu, \Ww_\nu] \right) \nonumber \\ &=&\frac{\tau^i}{2}
\left(\partial_\mu W^{i}_\nu-\partial_\nu W^{i}_\mu -g
\epsilon^{ijk}W_{\mu}^{j}W_{\nu}^{k} \right)
\eeqn

The  Goldstone bosons, $\omega^i$, within the built-in SU(2)
symmetry are assembled in a matrix $\Sigma$
\beqn
\Sigma=exp(\frac{i \omega^i \tau^i}{v}) \;\; ; v=246~GeV\;\;
\;\mbox{ and} \;\; {{\cal D}}_{\mu} \Sigma=\partial_\mu \Sigma +
\frac{i}{2} \left( g \Ww_{\mu} \Sigma - g'B_\mu \Sigma \tau_3
\right)\; ; \;  g'=e/c_W \nonumber \\
\eeqn

This leads to the gauge invariant mass term for the $W$ and $Z$

\beqn
\cl_M=\frac{v^2}{4} \tr(\cd^\mu \Sigma^\dagger \cd_\mu \Sigma)
\equiv - \frac{v^2}{4}  \tr\left( \V_\mu \V^\mu \right) \;\; \;\;
;  \;\; \V_\mu=\left( {{\cal D}}_{\mu} \Sigma \right)
\Sigma^{\dagger} \;\; ; \;\;M_W=\frac{g v}{2}
\eeqn

Incidentally, in the unitary gauge, $\V_\mu$ corresponds to the
triplet of the massive gauge bosons $W^\pm, Z$. Note also that at
the next-to-leading order in the chiral Lagrangian approach there
are genuine quartic couplings, however they only involve the
massive vector bosons, $WWWW, WWZZ, ZZZZ$. These quartic couplings
can not, unfortunately, be studied at LEP2. They are described, in
the $SU(2)_c$ limit,

\beqn
\label{NLOsu2}
\cl_{NLO}= \frac{L_1}{16 \pi^2} \left( \tr (\V_\mu \V^\mu)
\right)^2 +\frac{L_2}{16 \pi^2} \left( \tr (\V^\mu \V_\nu)
\right)^2
\eeqn

Photonic quartic operators appear first as next-to-next-to-leading
operators. Even by requiring $SU(2)_c$, \cviol and \pviol
conservation  there are quite a few quartic photonic operators. We
list them below and show the contribution of each to the quartic
Lorentz structures of interest, described earlier in
Eqs.~\ref{coup_wwgg}-~\ref{coup_zzzg}. The $\dots$ represent possible $4W,
4Z, WWZZ$ as well as Goldstones vertices. The $k_i^j$ parameterise
the strength of the anomalous coupling. By exploiting properties
of the trace of unitary $2\times 2$ matrices, other possible
combinations of operators can be expressed as combinations of the
operators given below.

\beqn
\frac{k_0^w}{\Lambda^2} \; g^2\; \tr (\W_{\mu \nu} \W^{\mu \nu})
\tr (\V^\alpha \V_\alpha) &\rightarrow& k_0^w \left( \right.
\;\zng \;+\;\frac{c_W}{s_W} \znz \;+\; \wng \;+\;
\frac{c_W}{s_W}\wnz \;+\; \dots \nonumber \\ &&
\\ \frac{k_c^w}{\Lambda^2} \;g^2\; \tr (\W_{\mu \nu} \W^{\mu \alpha}) \tr (\V^\nu
\V_\alpha) &\rightarrow& k_c^w \left( \right. \zcg \;+\;\frac{c_W}{s_W} \zcz \;+\; \wcg
\;+\; \frac{c_W}{s_W}\wcz \;+\; \dots \nonumber \\
&&
\\ \frac{k_1^w}{\Lambda^2} \;g^2\; \tr (\W_{\mu \nu}
\V^\alpha) \tr ( \W^{\mu \nu} \V_\alpha) &\rightarrow& k_1^w
\left( \right. \zng \;+\;\frac{c_W}{s_W} \znz \;+\; \wuz \;+\;
\dots
\\
\frac{k_2^w}{\Lambda^2} \;g^2\; \tr (\W_{\mu \nu} \V^\nu) \tr (
\W^{\mu \alpha} \V_\alpha) &\rightarrow& k_2^w \left( \right. \zcg
\;+\;\frac{c_W}{s_W} \zcz \;+\; \wdz \;+\; \dots
\\
\frac{k_3^w}{\Lambda^2} \;g^2\; \tr (\W_{\mu \nu} \V_\alpha) \tr (
\W^{\mu \alpha} \V^\nu) &\rightarrow& k_3^w \left( \right. \zcg
\;+\;\frac{c_W}{s_W} \zcz \;+\; \wtz \;+\; \dots
\eeqn

\beqn
\frac{k_0^b}{\Lambda^2} \;g'^2 \tr (\B_{\mu \nu} \B^{\mu \nu}) \tr
(\V^\alpha \V_\alpha) &\rightarrow& k_0^b \left( \right. \zng
\;-\;\frac{s_W}{c_W} \znz \;+\; \wng \;-\; \frac{s_W}{c_W}\wnz
\;+\; \dots \nonumber \\ &&
\\
\frac{k_c^b}{\Lambda^2} \;g'^2 \tr (\B_{\mu \nu} \B^{\mu \alpha})
\tr (\V^\nu \V_\alpha) &\rightarrow& k_c^b \left( \right. \zcg
\;-\;\frac{s_W}{c_W} \zcz \;+\; \wcg \;-\; \frac{s_W}{c_W}\wcz
\;+\; \dots \nonumber \\ &&
\\\frac{k_1^b}{\Lambda^2} \; g'^2 \tr (\B_{\mu \nu} \V^\alpha) \tr ( \B^{\mu \nu} \V_\alpha)
&\rightarrow& k_1^b \left( \right. \zng \;-\;\frac{s_W}{c_W} \znz \;+\; \dots
\\
\frac{k_2^b}{\Lambda^2} \;g'^2 \tr (\B_{\mu \nu} \V^\nu) \tr (
\B^{\mu \alpha} \V_\alpha) &\rightarrow& k_2^b \left( \right. \zcg
\;-\;\frac{s_W}{c_W} \zcz \;+\; \dots
\eeqn

and

\beqn
\frac{k_0^m}{\Lambda^2} \;g g' \tr (\W_{\mu \nu} \B^{\mu \nu}) \tr
(\V^\alpha \V_\alpha) &\rightarrow& k_0^m \left( \right. \zng
\;+\; c_{ZW} \znz \;+\; \wng \;+\;c_{ZW} \wnz \;+\; \dots
\nonumber \\ &&
\\ \frac{k_c^m}{\Lambda^2} \;g g' \tr (\W_{\mu \nu} \B^{\mu
\alpha}) \tr (\V^\nu \V_\alpha)&\rightarrow& k_c^m \left( \right.
\zcg \;+\;c_{ZW} \zcz \;+\; \wcg \;+\;c_{ZW} \wcz \;+\; \dots
\nonumber \\ &&
\\
\frac{k_1^m}{\Lambda^2} \;g g' \tr (\W_{\mu \nu} \V^\alpha) \tr (
\B^{\mu \nu} \V_\alpha) &\rightarrow& k_1^m \left( \right. \zng
\;+\;c_{ZW}
 \znz \;+\; \frac{1}{2}\wuz \;+\; \dots
\\
\frac{k_2^m}{\Lambda^2} \;g g' \tr (\W_{\mu \nu} \V^\nu) \tr (
\B^{\mu \alpha} \V_\alpha) &\rightarrow& k_2^m \left( \right. \zcg
\;+\;c_{ZW}
 \zcz \;+\; \frac{1}{2}\wdz \;+\; \dots
\\
\frac{k_3^m}{\Lambda^2} \;g g' \tr (\W_{\mu \nu} \V^\alpha) \tr (
\B^{\mu \alpha} \V_\nu ) &\rightarrow& k_3^m \left( \right. \zcg
\;+\;c_{ZW} \zcz \;+\; \frac{1}{2}\wtz \;+\; \dots
\eeqn
with
\beqn
c_{ZW} \equiv cotg2 \theta_W=\frac{c_W^2-s_W^2}{2 c_W s_W} \nonumber
\eeqn

There are a few observations that one can make. First, this
construction shows that the number of gauge invariant operators
exceeds the number of Lorentz structures,
Eqs.~\ref{coup_wwgg}~-~\ref{coup_zzzg}, which may be probed by the
three processes, $\epem \ra W^+W^-\gamma, ZZ\gamma,
Z\gamma\gamma$. Note that the $k_{1,2,3}^{w,b,m}$ do not
contribute to \ggwwt and therefore have no connection to the
operators $a_{0,c}$ that were introduced in
\cite{nousggvv,nousee3v}. In fact $k_{1,2}^b$ does not even
contribute to \eewwgt. Note also that a limit on $k_{1,2,3}^w$
from the process \eewwgt can be directly translated as a limit on
$k_{1,2,3}^m/2$.

We see that contrary to the claim in
\cite{Stirlingquartic,eboliquartic}, the fact that we have used a
manifestly gauge invariant and $SU(2)_c$ symmetric approach shows
that operators which contribute to $WW\gamma\gamma$,
$k_{0,c}^{w,b,m}$, do in general induce a $WWZ\gamma$ vertex. In
\cite{nousee3v} only the structures $\gamma \gamma VV$, $V=W,Z$,
were considered in order to compare with limits extracted from the
laser mode of the LC \cite{nousggvv}. Therefore strictly speaking
the analysis in \cite{nousee3v} assumes a relation between the
$k_i$ such that the $WWZ\gamma$ (and also the $ZZZ\gamma$)
vanishes. With $k_{1,2,3}^{w,b,m}=0$, the general condition for
the vanishing of the $WWZ\gamma$ and $ZZZ\gamma$ vertices is
$2k^w_{0,c}+k^m_{0,c}= 2 \sin^2 \theta_W
(k^b_{0,c}+k^w_{0,c}+k^m_{0,c})$, with
$k^b_{0,c}+k^w_{0,c}+k^m_{0,c} \neq 0$, so that one does not also
make the $VV\gamma \gamma$ vanish. One very simple implementation
of this condition is to have, all $k_i=0$ apart from $k_{0,c}^w$
and $k_{0,c}^b$ with the constraint
\beqn
\label{nowwz}
k_{0,c}^w&=&k_{0,c}^{\gamma \gamma}\; s_W^2 \nonumber \\
k_{0,c}^b&=&k_{0,c}^{\gamma \gamma}\; c_W^2 \nonumber \\
\eeqn
 we then end up with only two independent parameters
controlling $WW\gamma\gamma$ like in the analysis in
\cite{nousggvv}. With the constraint on the vanishing of
$WWZ\gamma$, we can  make contact with the original operators
introduced in \cite{nousggvv,nousee3v}. We then have, with the
constraint Eq.~\ref{nowwz}
\beqn
\label{coeffa0}
 a_{0,c}=4 g^2 (k_{0,c}^w + k_{0,c}^b + k_{0,c}^m )=4 g^2 k_{0,c}^{\gamma \gamma}
\eeqn

%

On the other hand we can arrange the operators such that the
$SU(2)_c$ $\gamma \gamma VV$ couplings vanish so that effectively
$a_{0,c}=0$, but not the quartic $WWZ\gamma$. For instance with
all $k_i=0$ but $k_{0,c}^b, k_{0,c}^w$, this can be achieved by
having $k_{0,c}^b = -k_{0,c}^w$.

In the basis, for the chiral Lagrangian, that we have chosen all
operators are seen to contribute to $\epem \ra Z\gamma\gamma,
ZZ\gamma$. However it is easy to see that we can choose
combinations of $k_i^j$ such that all $ZZ\gamma\gamma$ and
$ZZZ\gamma$ vanish, in which case only \eewwgt will provide a test
on the quartic photonic anomalous couplings. For example taking
$k_2^m=-k_3^m$, with all other parameters set to zero,  only
leaves the $WWZ\gamma$ vertex. Also because all the operators map
on only two distinct $ZZ\gamma\gamma$ Lorentz structures, $\epem
\ra Z\gamma \gamma$ can not discriminate between the various
operators.

The argument in \cite{Stirlingquartic} that there can not be a
dim-4 (in the U-gauge) operator with photons because of custodial
symmetry is incorrect. The reason is $U(1)$ gauge invariance as
stated in \cite{nousee3v}. The authors of
\cite{eboliquartic,Stirlingquartic} consider another operator
which contributes to $WWZ\gamma$ but not to $WW\gamma \gamma$.
Though that operator can be made $SU(2)_c$ invariant it explicitly
breaks \cpviol (see Appendix) , and therefore we do not consider
it here nor do we consider any of the quartic couplings that
violate other discrete symmetries.

\section{Linear approach to embedding the photonic quartic couplings}
As shown repeatedly, see for instance \cite{Morioka,Cliff}, any operator can be made
gauge invariant even in the linear approach with the presence of a Higgs. What changes is
the hierarchy in the couplings. For instance the equivalent of the structures $k_{0,c}$
are
\beqn
{{\cal L}}_{0}&=&\frac{1}{\Lambda^4} (D_\mu \Phi) (D^\mu
\Phi)^\dagger \times \;\;\left\{g'^2\; Q_0^b \;B_{\alpha \beta}
B^{\alpha \beta}\;\;+\;\; g^2 \;Q_0^w \;W_{\alpha \beta}^i
W^{i\;\alpha \beta} \;\;+\;\;g g'\; Q_0^m\; W^{3}_{\alpha \beta}
B^{\alpha \beta} \right\} \nonumber \\
\eeqn

and

\beqn
{{\cal L}}_{c}&=&\frac{1}{\Lambda^4} \;\; \frac{1}{2} \left\{ (D^\mu \Phi) (D_\nu
\Phi)^\dagger +h.c. \right\} \times \nonumber \\
&& \left\{g'^2\; Q_c^b \;B_{\alpha \mu }
B^{\alpha \nu}\;\;+\;\; g^2 \;Q_c^w \;W_{\;\alpha \mu }^i W^{i\;\alpha \nu } \;\;+\;\;g
g'\; Q_c^m\; W^{3}_{\alpha \mu } B^{\alpha \nu } \right\}
\eeqn

Written in terms of the fundamental fields of the \sm the above
operators lead to quartic couplings but also to vertices with up
to 8 legs! All the operators  contribute to $\gamma \gamma
WW,\gamma \gamma ZZ, \gamma Z WW,\gamma ZZZ, ZZWW,ZZZZ$, while
$Q^w_{0,c}$ contributes also to $WWWW$.


When studying the quartic anomalous couplings, we can make the
following equivalence
\beqn
\frac{Q_i^j}{\Lambda^2} =- \frac{1}{2} \frac{g^2}{M_W^2} k_i^j
\;\;\; ; \;\;\; i=0,c
\eeqn

However the main difference is that in the linear approach {\it
\`a la } \sm, the operators are dimension $8$ operators. Therefore
to be consistent one should also list operators of the form
$F_{\mu \nu}^4$ which lead to $4\gamma$. This shows once more that
quartic operators are more likely in the event that there is no
Higgs. This observation has already been made for the leading
order $WWWW,WWZZ,ZZZZ$ operators \cite{Morioka}. The equivalent
operators corresponding to $k_{1,2,3}^{w,b,m}$ can also be easily
written within the linear approach, but we refrain from doing so.

\section{Analysis}

The computation of the different cross sections have been checked
at different levels by comparing the outputs of a hand calculation
implemented in the program used in \cite{nousee3v} against those
of two automatic programs for the generation of Feynman diagrams
and calculations of cross sections: {\tt GRACE} \cite{Grace} and
{\tt CompHEP} \cite{Comphep}. The former enables checks of the
polarised cross sections. Moreover, with {\tt CompHEP} all the new
operators, even in their explicit $SU(2)\times U(1)$ forms, have
been implemented at the Lagrangian level through an interface with
{\tt LANHEP} \cite{Lanhep}. The latter, given the Lagrangian,
automatically
 generates  all the Feynman rules and
vertices in a format which is read directly by {\tt CompHEP},
therefore one can say that the checks have been performed even at
the level of the Feynman rules, thanks to {\tt LANHEP}
\cite{Lanhep}.

In all our calculations we have taken: $M_Z=91.18~GeV,
M_W=80.41~GeV, \sin^2\theta_W=1-M_W^2/M_Z^2$ and
$\alpha(M_Z)^{-1}=128.07$. However the electromagnetic coupling
involving any external photon is set to $\alpha=1/137.035$. When
stating limits on the anomalous couplings we will take
$\Lambda=M_W$, all limits can be trivially rescaled for any other
choice of $\Lambda$. All our analysis is based on the total
$WW\gamma$ and $ZZ\gamma$ cross section allowing for all decay
products of the $W$ and $Z$. We have not considered the added
effect of any anomalous tri-linear coupling, as already stressed
the latter are much better probed in $\epem \ra W^+ W^-$. In an
experimental setting, the signature to consider is the one with
$4$-fermions and an energetic photon. There are then other
contributions, which depend on the 4-f final state, which are not
mediated through the diagrams that contribute to $WW\gamma$ with
both $W$ decaying. The full 4-fermion $+\gamma$ contributions have
been thoroughly studied very recently \cite{4fgamma}. These
background contributions, not going through the resonant
$WW\gamma$ contribution, are less important at LEP2 energies
\cite{4fgamma}. Moreover invariant mass cuts such that the
4-fermions reconstruct a $W$ pair and are central (to reduce
``single W" production), should drastically suppress these
background contributions.

Already at this stage we can guess the main characteristics of the
distributions. The use of the field tensor for the photon means
that the anomalous terms lead energetic photons which will be
preferentially produced in the central region, contrary to the \sm
photons which are essentially radiative bremsstrahlung photons. For
LEP2 we confine our analysis to the ultimate LEP2 energy of
200~GeV and assume a luminosity of $150$~pb$^{-1}$. For higher
energies  to illustrate how drastic the
improvement is, we take $\sqrt{s}=500$~GeV and ${{\cal
L}}=500$~fb$^{-1}$.
\subsection{\eewwgt at LEP2}
\begin{figure*}[thbp]
\begin{center}
\mbox{
\mbox{\epsfxsize=8.5cm\epsfysize=10cm\epsffile{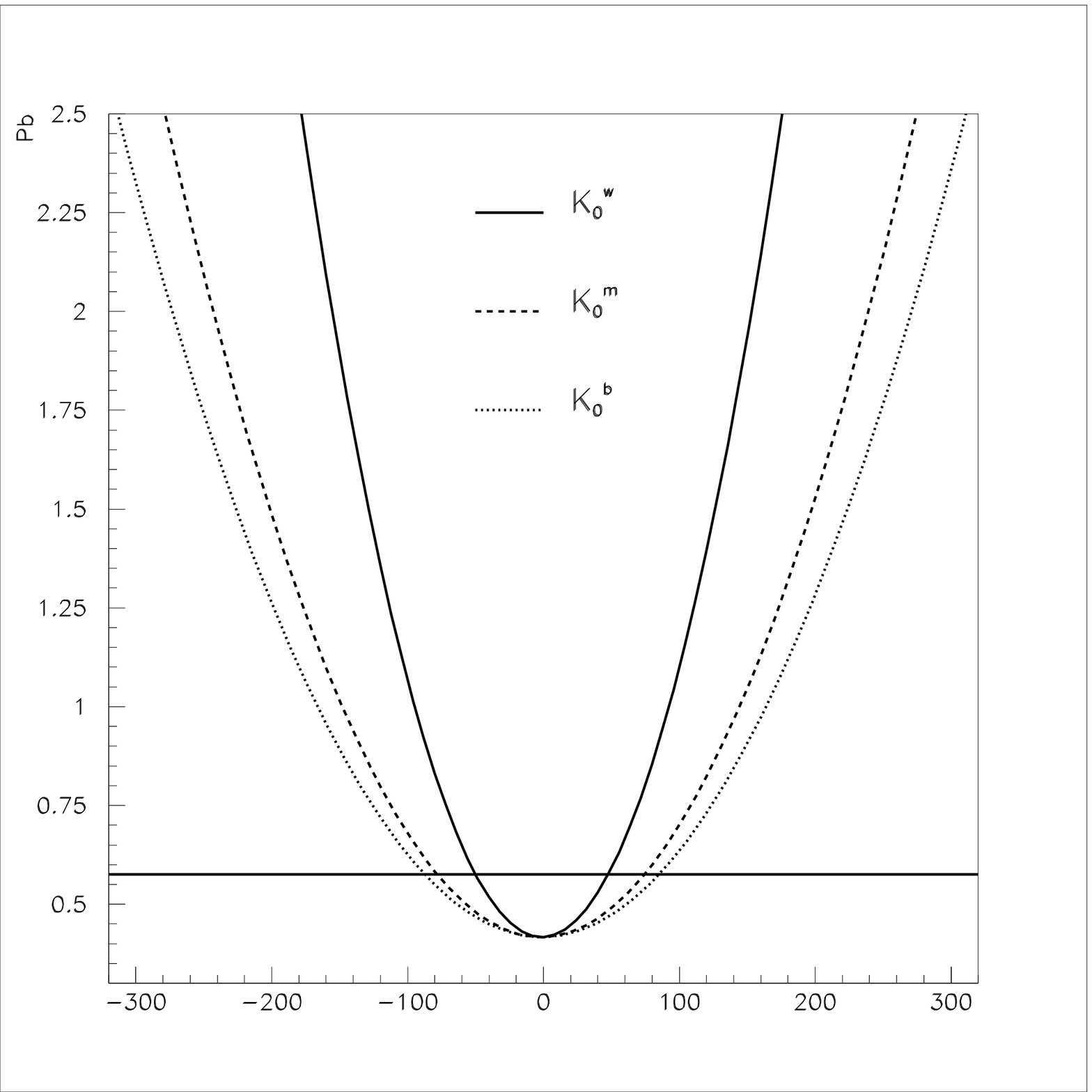}}
\mbox{\epsfxsize=8.5cm\epsfysize=10cm\epsffile{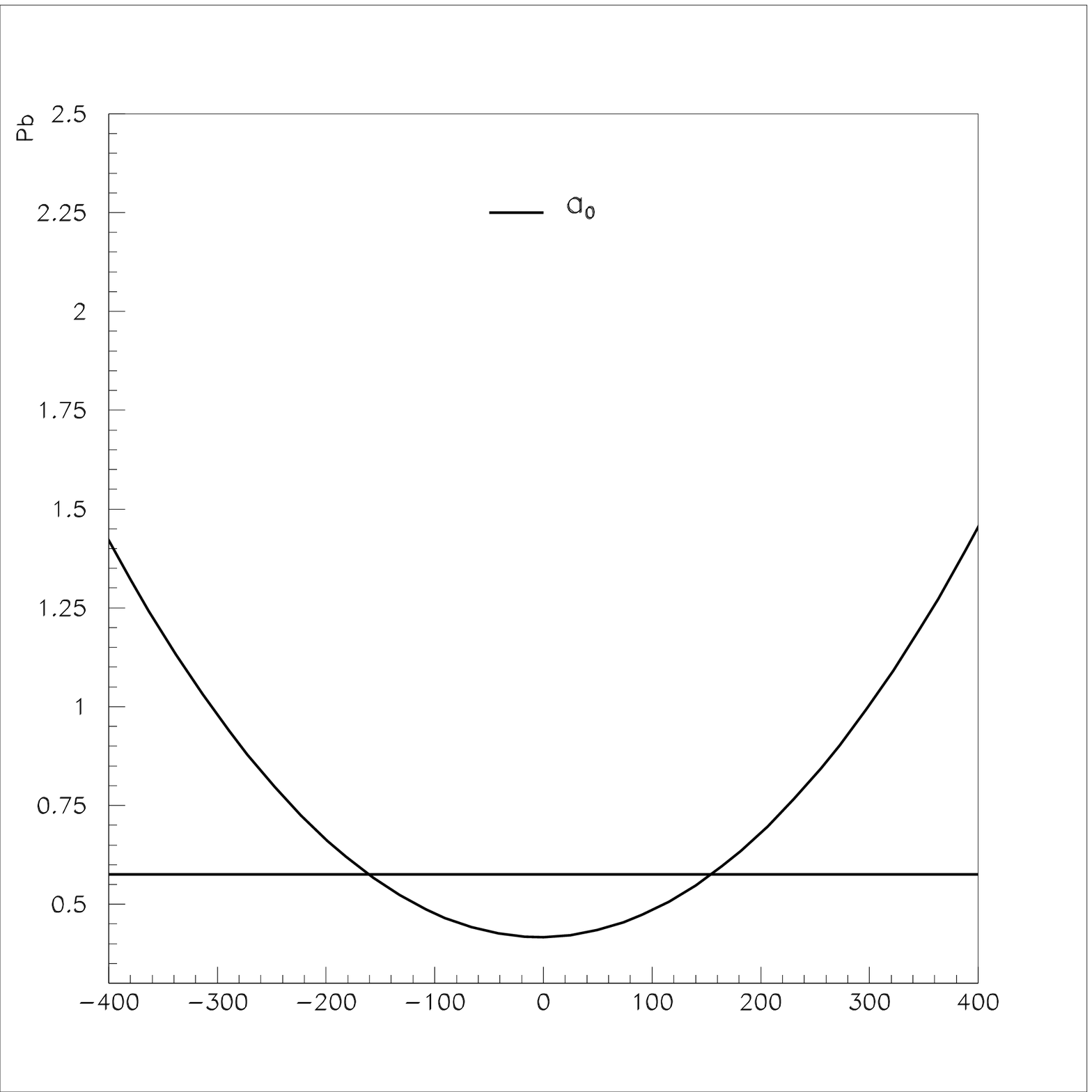}} }
\vspace*{1cm}
\mbox{
\mbox{\epsfxsize=8.5cm\epsfysize=10cm\epsffile{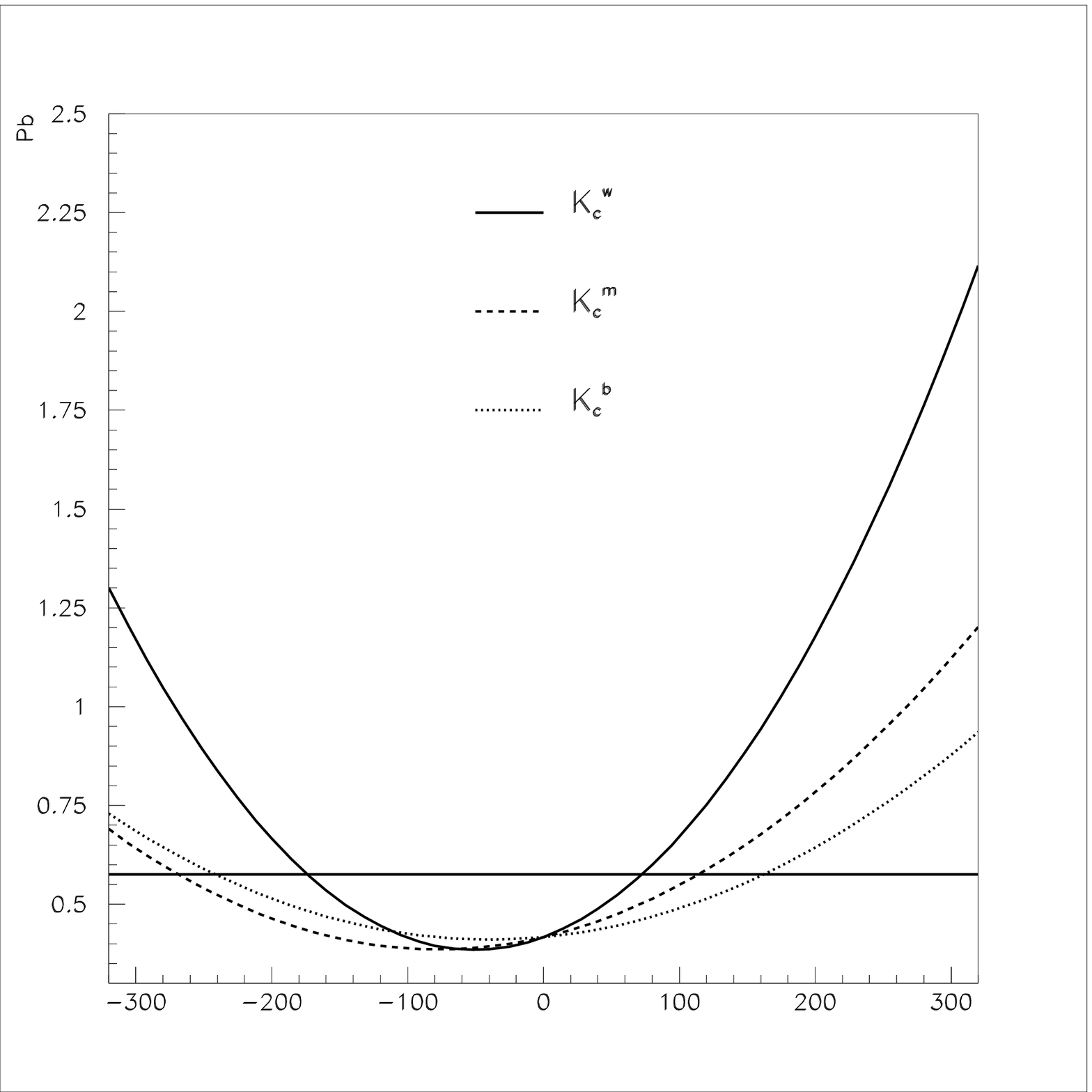}}
\mbox{\epsfxsize=8.5cm\epsfysize=10cm\epsffile{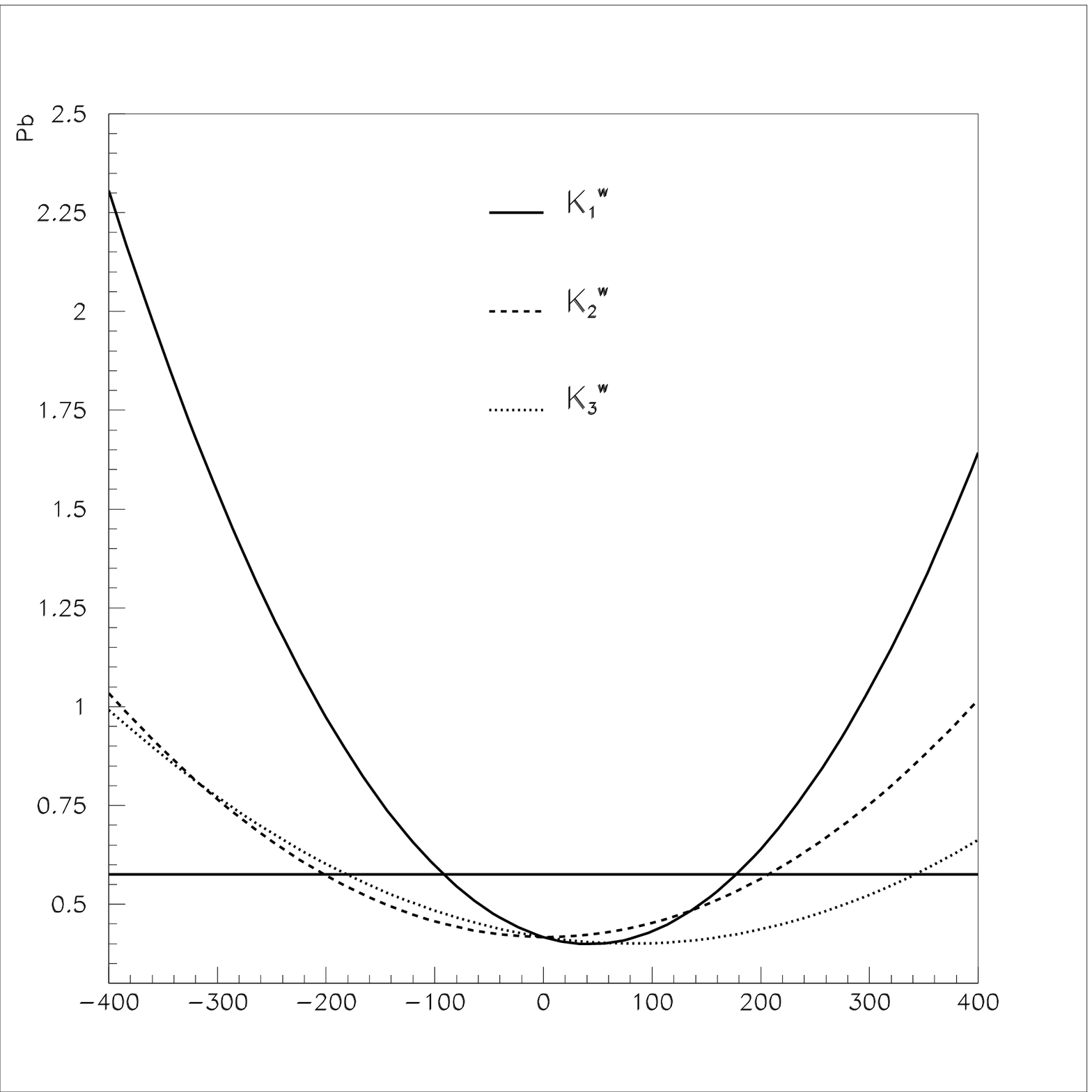}} }
\caption{\label{ki-dependence-lep2}{\em Dependence of the \eewwgt cross section
at $\sqrt{s}=200$~GeV on the anomalous parameters a) $k_0$, b)
$a_0$ with the constraint ~Eq.~(\ref{nowwz})  c) $k_c$ and d)
$k_{1,2,3}^w$. For the cuts on the photon refer to the text. The horizontal line
indicates the $3\sigma$ increase of the cross section. $\Lambda$ has been set to $M_W$.} }
\end{center}
\end{figure*}

Our cuts on the photon energy, $E_\gamma$, and its angle with the
beam, $\theta_\gamma$, are such that  $E_\gamma>5~GeV$ and
$20^0<\theta_\gamma<160^0$. The \sm cross section is then
$.417$~pb. With the design luminosity of $150$~pb$^{-1}$ this
amounts to about $60$ events, before any efficiency or selection
factor is included. Fig.~\ref{ki-dependence-lep2} shows the
dependence of the total cross section, at $\sqrt{s}=200$~GeV, on
the parameters $k_i$. As discussed in the previous section,
$k_{1,2}^b$ do not contribute to \eewwgt, moreover the limits one
extracts from $k_{1,2,3}^w$ can be directly translated to
$k_{1,2,3}^m$ (there is only  a factor 2 to apply between the
limits) since both only contribute to the $WWZ\gamma$ coupling. On
the other hand in our classification, this is not true for the
$k_{0,c}^{w,b,m}$ since each gives a different weight to the
$WWZ\gamma$ coupling compared to the $WW\gamma \gamma$ and
therefore we show all of the $k_{0,c}$ dependencies. As an
illustration we also show a model with the constraints
Eq.~(\ref{nowwz}) where only the anomalous $WW\gamma \gamma$
coupling survives and hence is amenable to a description in terms
of $a_0$. One notices that the $k_0^i$ (including $a_0$) interfere
very little with the \sm compared to the other couplings. As we
will see, this explains why two values of $k_i, i\neq 0$ which
give the same cross section can give markedly different
distributions. From these figures  with the simple cuts that we
have assumed, a $3\sigma$ measurement of the cross section allows
to set the following individual limits:

\begin{figure*}[bhtp]
\begin{center}
\mbox{
\mbox{\epsfxsize=8.cm\epsfysize=10cm\epsffile{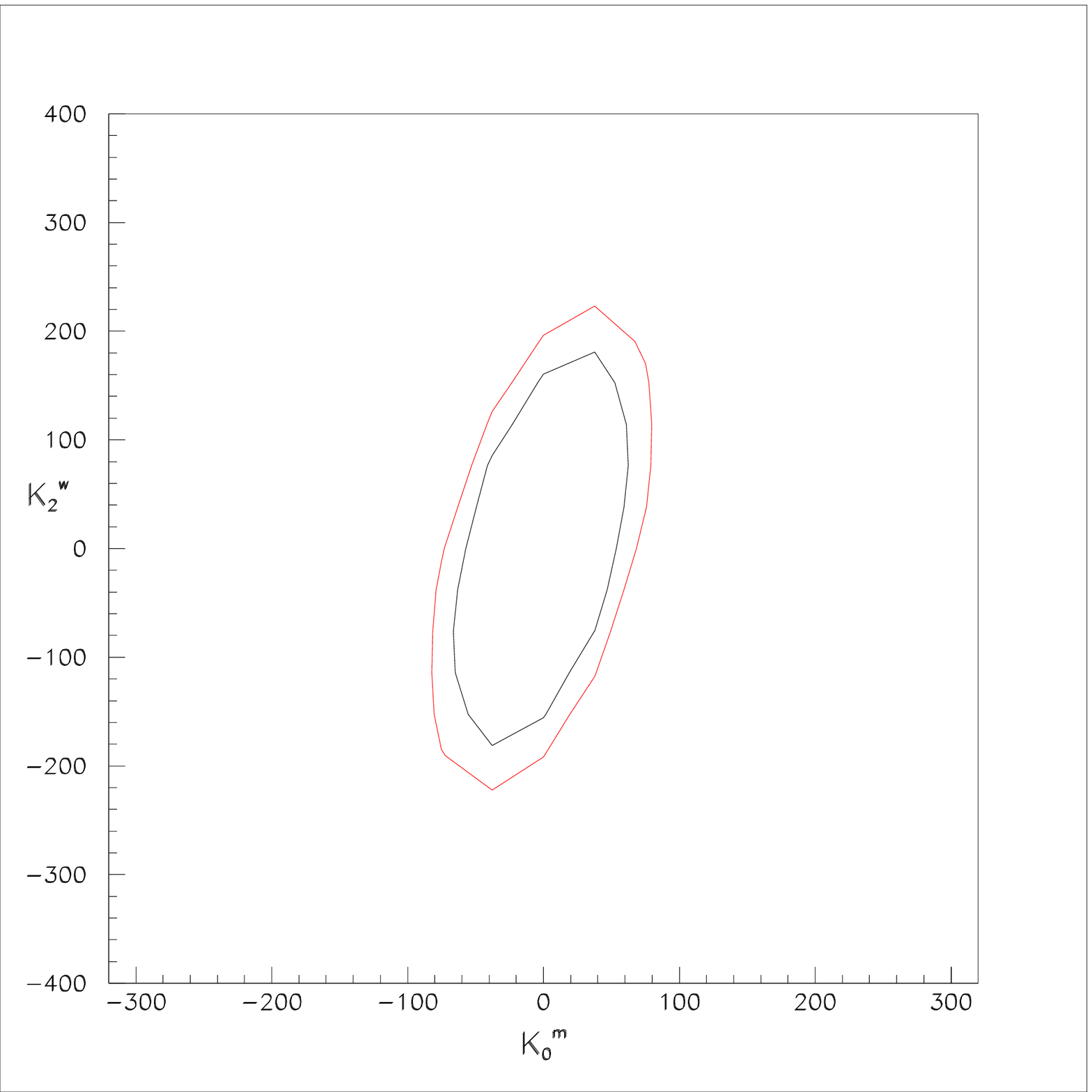}}
\mbox{\epsfxsize=8.cm\epsfysize=10cm\epsffile{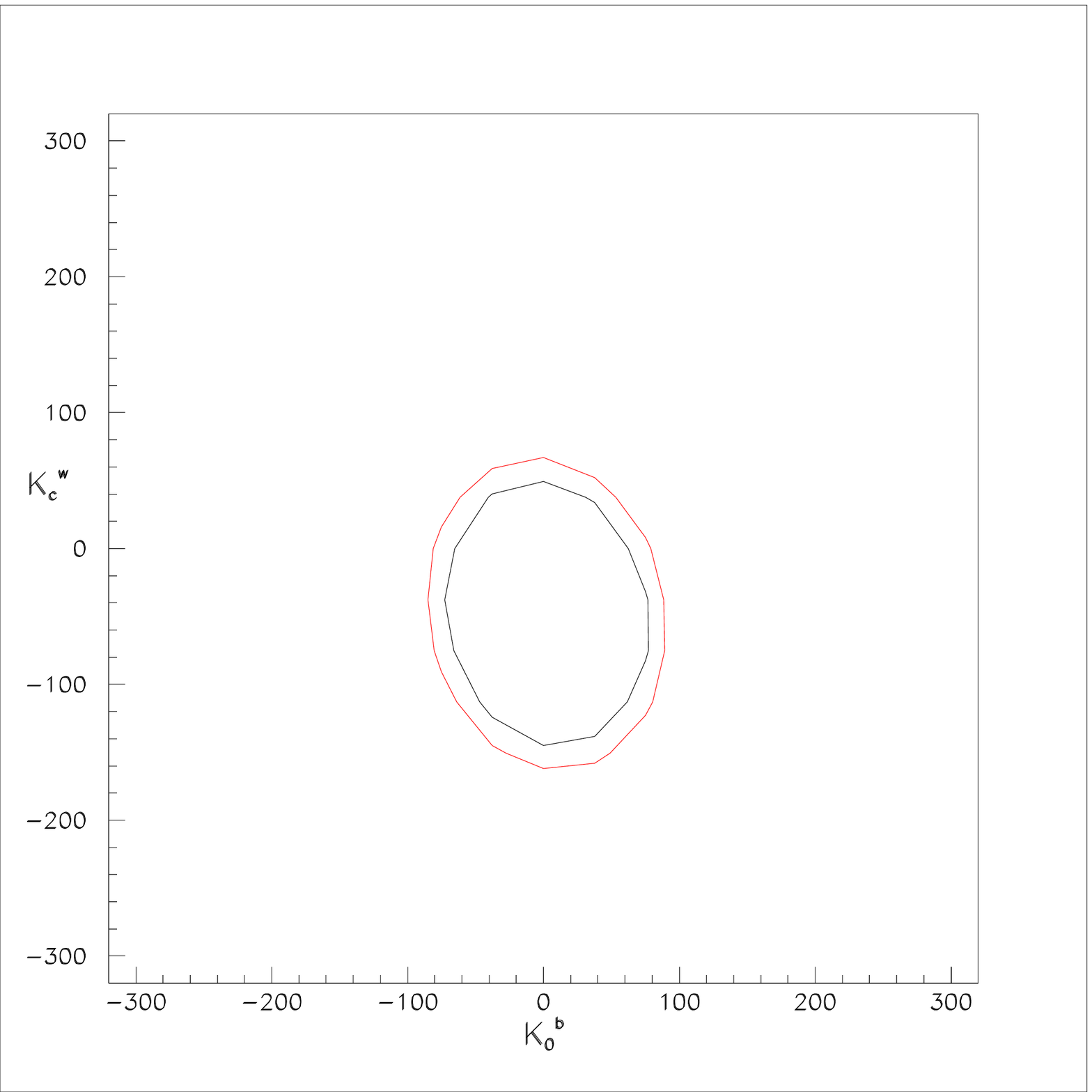}} }
\caption{\label{contours-ki}{\em $2\sigma$ and $3\sigma$ contours
from \eewwgt at $\sqrt{s}=200$~GeV in $k_0^m-k_2^w$ and the $k_0^b-k_c^w$
planes.
}}
\end{center}
\end{figure*}
\beqn
\label{eewwglep2limit}
-1.35\; 10^{-2}\;{\rm GeV}^{-2}&<&\frac{k_0^b}{\Lambda^2}<1.30\;
10^{-2} \;{\rm GeV}^{-2}  \nonumber
\\ -0.78\; 10^{-2}\; {\rm
GeV}^{-2}&<&\frac{k_0^w}{\Lambda^2}<0.73\; 10^{-2} \;  {\rm
GeV}^{-2}  \nonumber
\\ -1.20 \;10^{-2}\; {\rm
GeV}^{-2}&<&\frac{k_0^m}{\Lambda^2}<1.14\; 10^{-2}\;{\rm GeV}^{-2}
\nonumber
\\ -3.70 \;10^{-2} \; {\rm
GeV}^{-2}&<&\frac{k_c^b}{\Lambda^2}<2.50\; 10^{-2}\; {\rm
GeV}^{-2} \nonumber
\\ -2.69 \;10^{-2}\; {\rm
GeV}^{-2}&<&\frac{k_c^w}{\Lambda^2}<1.13\; 10^{-2}\; {\rm
GeV}^{-2} \nonumber
\\ -4.14 \;10^{-2}\; {\rm
GeV}^{-2}&<&\frac{k_c^m}{\Lambda^2}<1.77 \;10^{-2}\; {\rm
GeV}^{-2} \nonumber
\\ -1.44\; 10^{-2}\;{\rm
GeV}^{-2}&<&\frac{k_1^w}{\Lambda^2},\frac{k_1^m}{2\Lambda^2}
<2.73\; 10^{-2} \;{\rm GeV}^{-2}  \nonumber
\\ -3.10\; 10^{-2}\;
{\rm
GeV}^{-2}&<&\frac{k_2^w}{\Lambda^2},\frac{k_2^m}{2\Lambda^2}<3.20\;
10^{-2} \;  {\rm GeV}^{-2}  \nonumber
\\ -2.82 \;10^{-2}\; {\rm
GeV}^{-2}&<&\frac{k_3^w}{\Lambda^2},\frac{k_3^m}{2\Lambda^2}<5.28\;
10^{-2}\;{\rm GeV}^{-2}\nonumber
\\ -2.48 \;10^{-2}\; {\rm
GeV}^{-2}&<&\frac{a_0}{\Lambda^2}<2.39 \;10^{-2}\; {\rm GeV}^{-2}
\leftrightarrow -1.59 \;10^{-2}\; {\rm GeV}^{-2}<\frac{k_0^{\gamma
\gamma}}{\Lambda^2}<1.49 \;10^{-2}\; {\rm GeV}^{-2} \nonumber \\
\eeqn

We have also considered  correlations for
some specific combinations of couplings. As an illustration we
show the correlations in  the $k_0^m-k_2^w$ and the $k_0^b-k_c^w$
planes, see Fig.~\ref{contours-ki}. In each instance all the other couplings are set to zero.

\begin{figure*}[thp]
\begin{center}
\mbox{
\mbox{\epsfxsize=8.5cm\epsfysize=10cm\epsffile{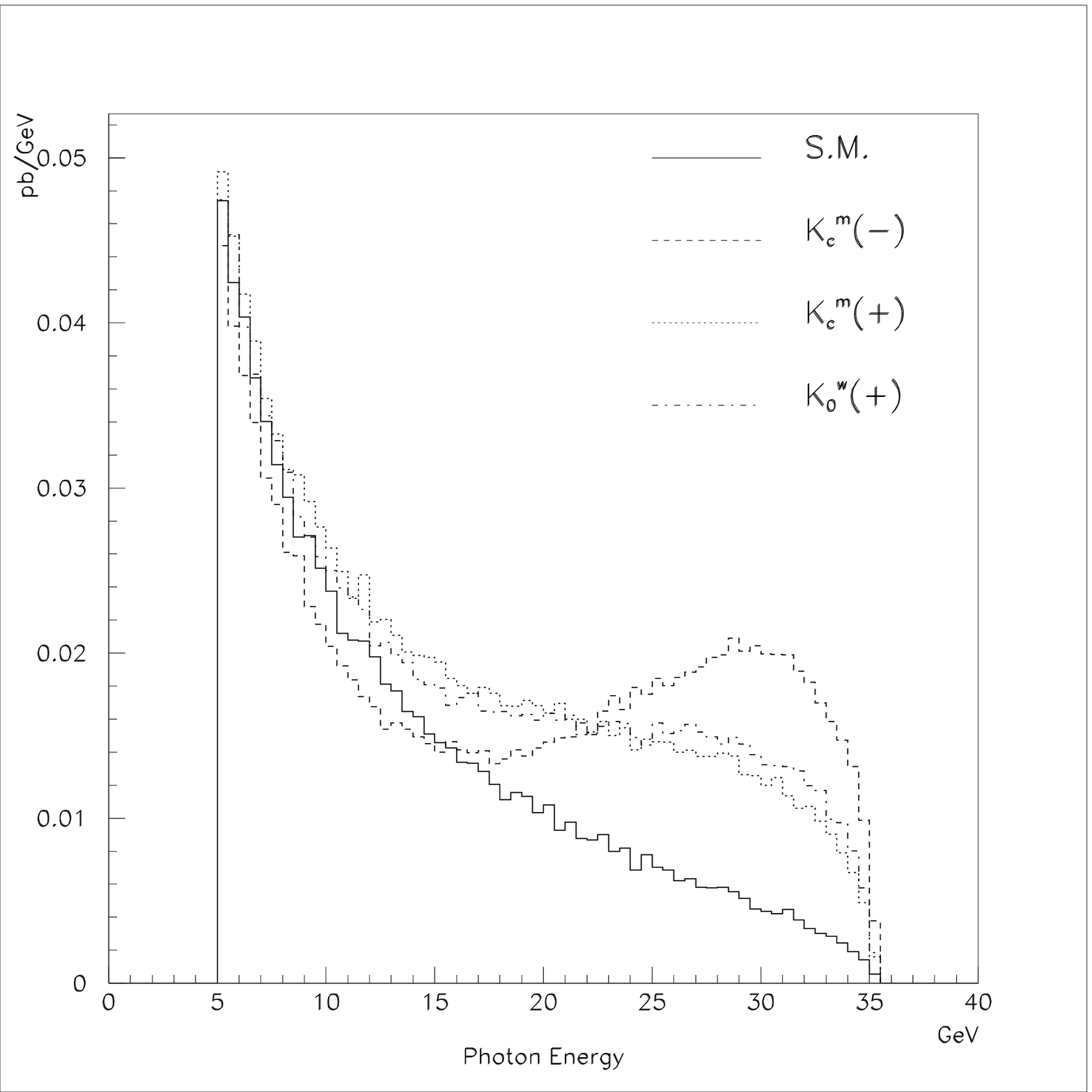}}
\mbox{\epsfxsize=8.5cm\epsfysize=10cm\epsffile{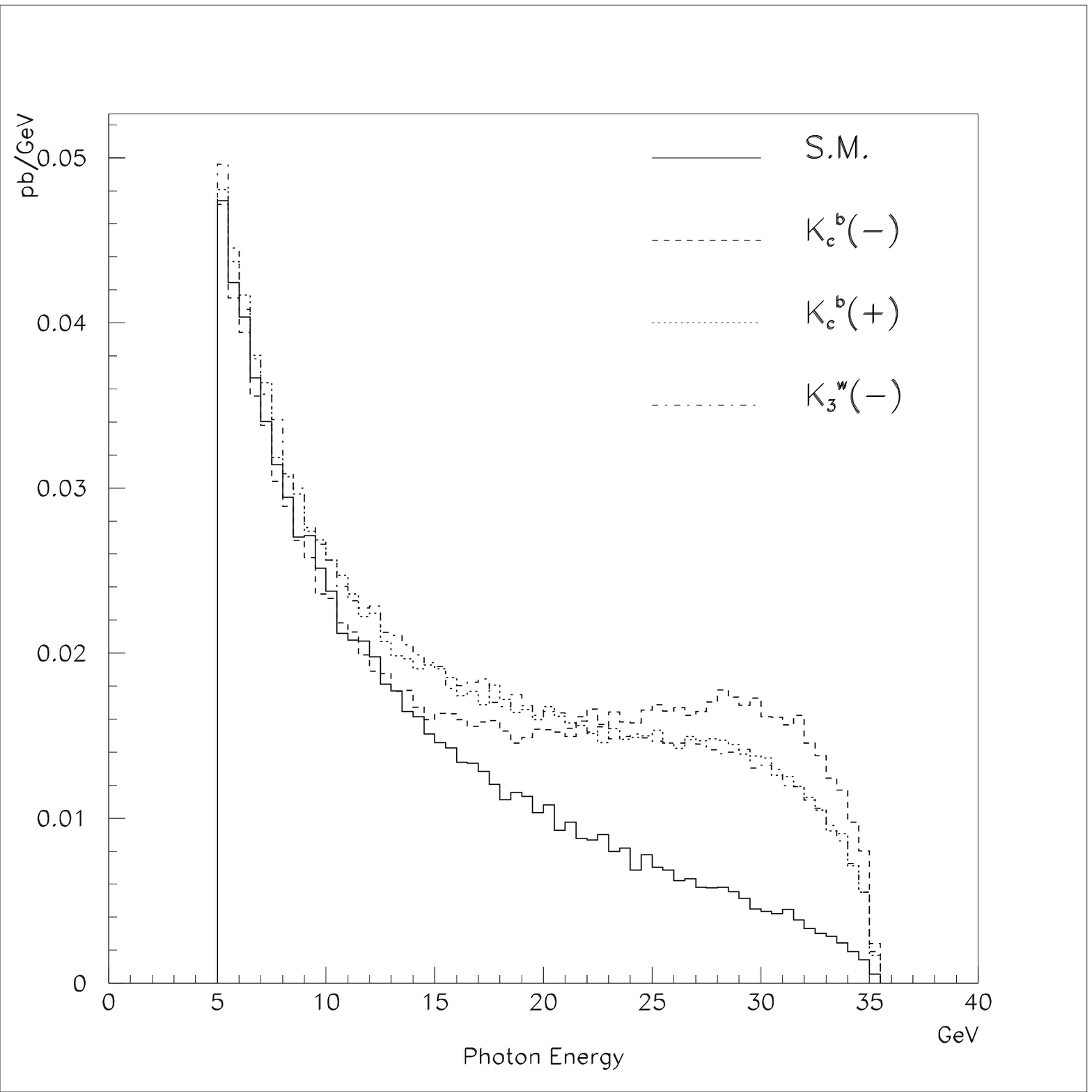}} }
\caption{\label{dis_eg}{\em Distribution in the energy of the photon in
\eewwgt due to the anomalous couplings
$k_0^b $ and $k_c^b$ as compared to the tree-level \sm. We have taken $k_i$ values that lead to a
$3\sigma$ increase of the cross section as explained in the text.
The $\pm$ are the positive and negative values given by
Eq.~(\ref{eewwglep2limit}).
}}
\end{center}
\end{figure*}

\begin{figure*}[htbp]
\begin{center}
\mbox{
\mbox{\epsfxsize=8.5cm\epsfysize=10cm\epsffile{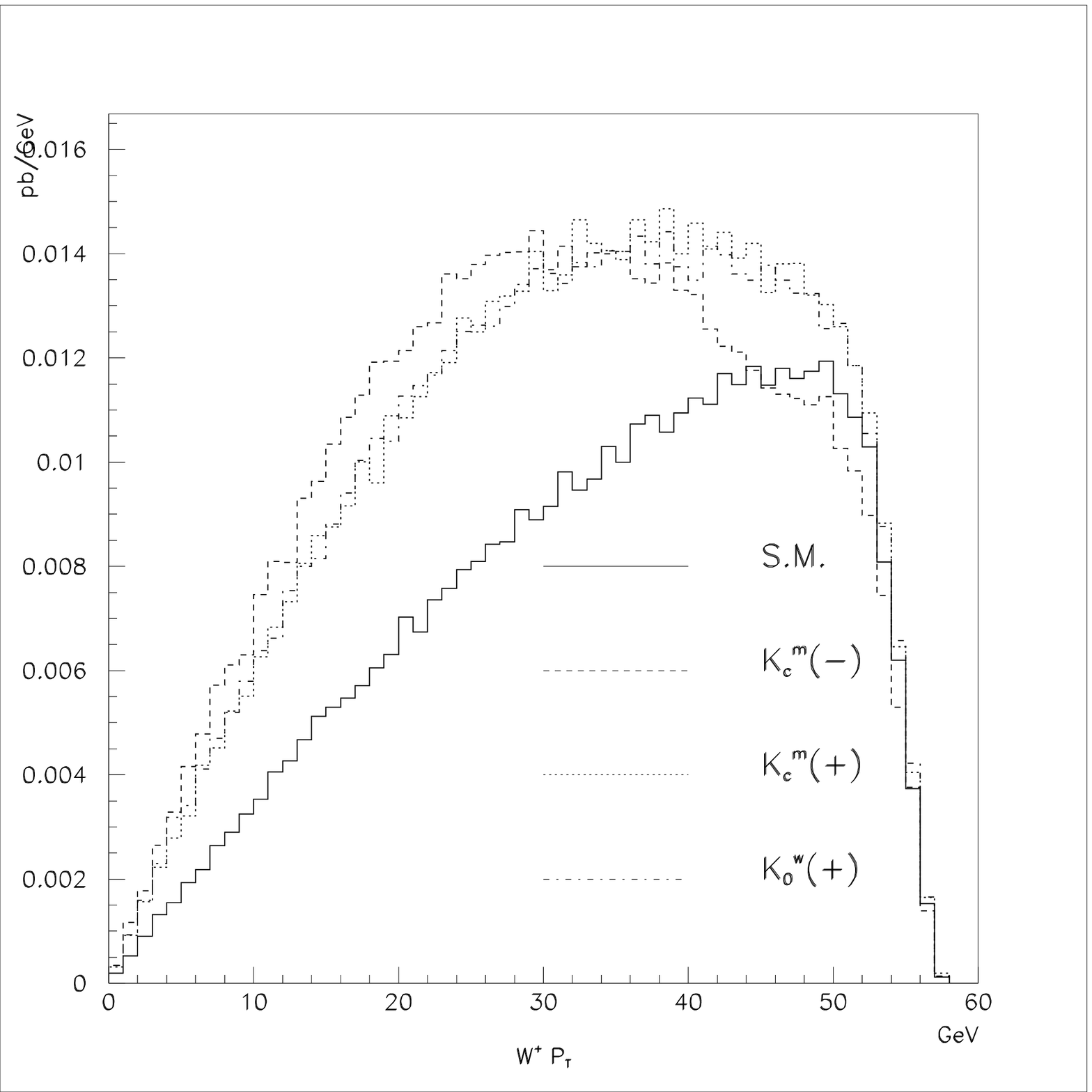}}
\mbox{\epsfxsize=8.5cm\epsfysize=10cm\epsffile{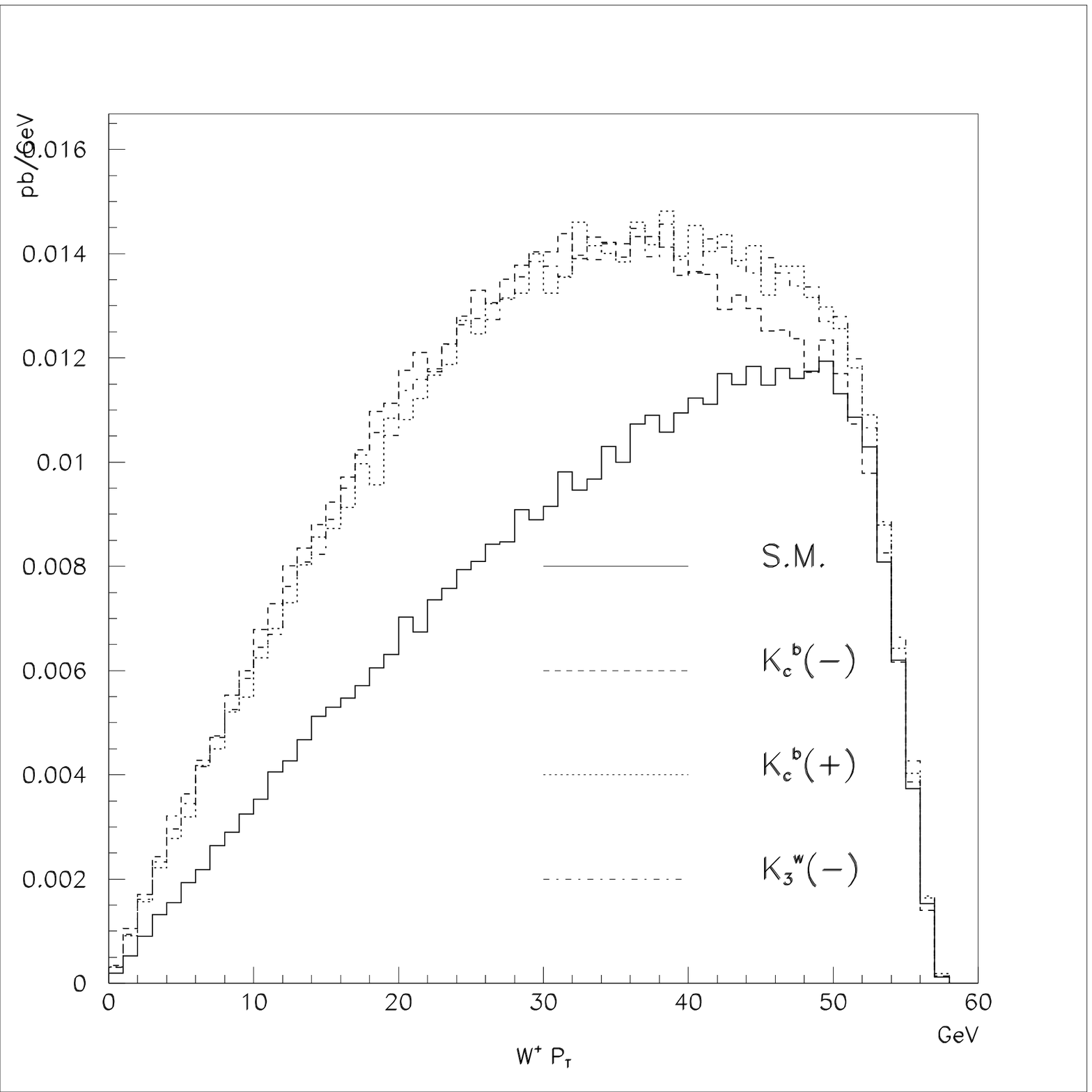}} }
\caption{\label{wwgptw}{\em As in Fig.~\ref{dis_eg} but for the $p_T$ of one of the
$W$'s.
}}
\end{center}
\end{figure*}

We now turn to the distributions. As explained above we expect the
distribution in the photon energy to be most revealing. This is
borne out by our analysis where we do find that these couplings lead to energetic
photons. To compare the various couplings we
have chosen all $k_i$ such that they all give  a $3\sigma$
increase in the \eewwgt cross section at $\sqrt{s}=200$~GeV with ${{\cal L}}=150$~fb$^{-1}$.
First, we note that all $k_0^i$ give the same distribution. This is
easily understood since we have found that these couplings interfered very
little with the \sm contribution and also because they all have
the same Lorentz structure\footnote{If one had polarised beams one could
exploit the fact that their $WWZ\gamma$ (mediated through a $Z$) and $WW\gamma \gamma$
(mediated through a photon) components are different to discriminate between them.}. However
this is not the case for the other couplings. For this class of
operators ($k_i^j \neq k_0^j$), we can, on the basis of the photon distribution
discriminate, between the two signs of the couplings of a same
operator (an effect
of the interference with the \sm) beside being able, in general, to differentiate between
different operators, see Fig~\ref{dis_eg}.  Another
interesting distribution to look at is the $p_T$ of the W, see
Fig~\ref{wwgptw}.
\subsection{$\epem \ra Z\gamma \gamma$ at LEP2}
\begin{figure*}[htbp]
\begin{center}
\mbox{\epsfxsize=12cm\epsfysize=12cm\epsffile{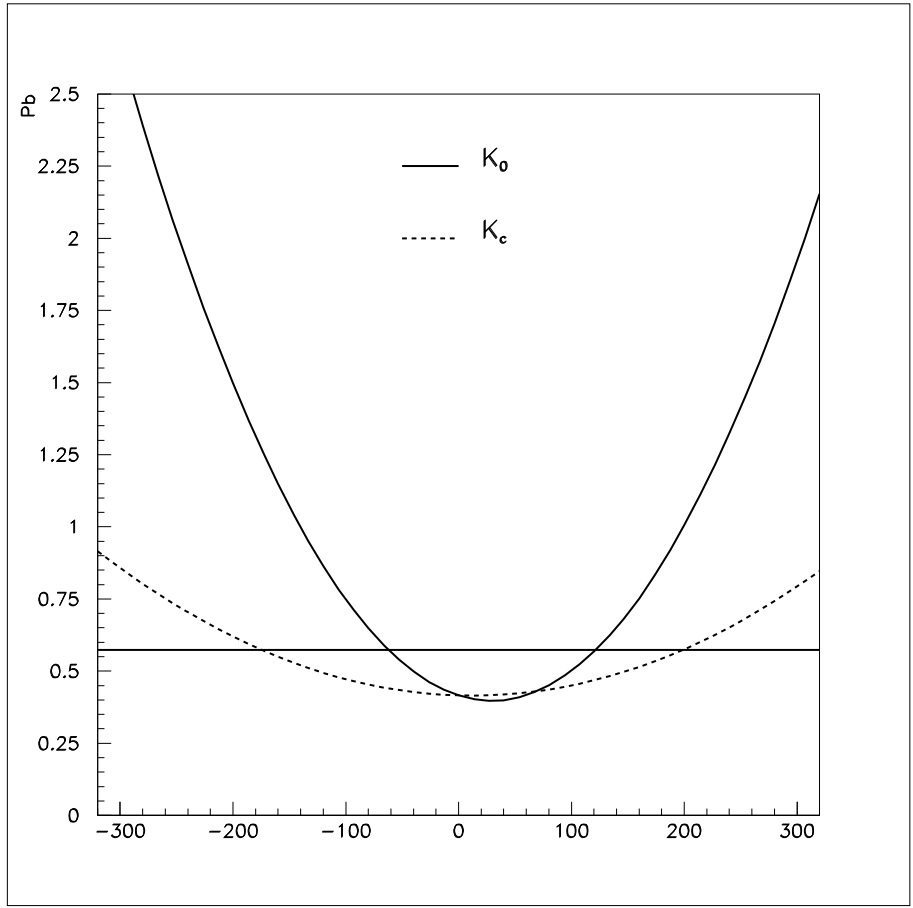}}
\caption{\label{ki-dep-aaz}{\em Dependence of the $\epem \ra Z\gamma \gamma$ cross section
at $\sqrt{s}=200$~GeV on the anomalous parameters $k_{0,c}$ with $\Lambda=M_W$. For the cuts on
the photon refer to the text.
}}
\end{center}
\end{figure*}

We take the same cuts on both photons as previously for \eewwgt.
We then have a cross section which is sensibly the same as the one
for  \eewwgt: $.416$~pb. As explained above we basically are
probing only two parameters $k_{0,1}^i$ and $k_{c,2,3}^j$. We have
chosen to show $k_0^b$ and $k_c^b$ dependencies in
Fig.~\ref{ki-dep-aaz}. In $Z\gamma\gamma$, the $k_0$ couplings
interfere more with the \sm contributions than in \eewwgt.

 For
the $3\sigma$ deviations we extract
\beqn
\label{limaazlep2}
-0.95\; 10^{-2}\;{\rm
GeV}^{-2}&<&\frac{k_0^b,k_0^w,k_0^m,k_1^b,k_1^w,k_1^m}{\Lambda^2}<1.87\;
10^{-2} \;{\rm GeV}^{-2}  \nonumber \\ -2.68\; 10^{-2}\; {\rm
GeV}^{-2}&<&\frac{k_c^b,k_c^w,k_c^m,k_2^b,k_2^w,k_2^m,k_3^w,k_3^m}{\Lambda^2}<3.07\;
10^{-2} \;  {\rm GeV}^{-2}  \nonumber \\ -1.48 \;10^{-2}\; {\rm
GeV}^{-2}&<&\frac{a_0}{\Lambda^2}<3.01 \;10^{-2}\; {\rm GeV}^{-2}
\eeqn
We now get limits on all couplings including $k_{1,2}^b$ which
were not probed in \eewwgt. What is more interesting is that
$\epem \ra Z\gamma \gamma$ sets slightly better limits on
$k_{2,3}^{w,m}$. For the other couplings, combining both reactions
improves the limits set by each process.

\begin{figure*}[htbp]
\begin{center}
\mbox{
\mbox{\epsfxsize=8.5cm\epsfysize=10cm\epsffile{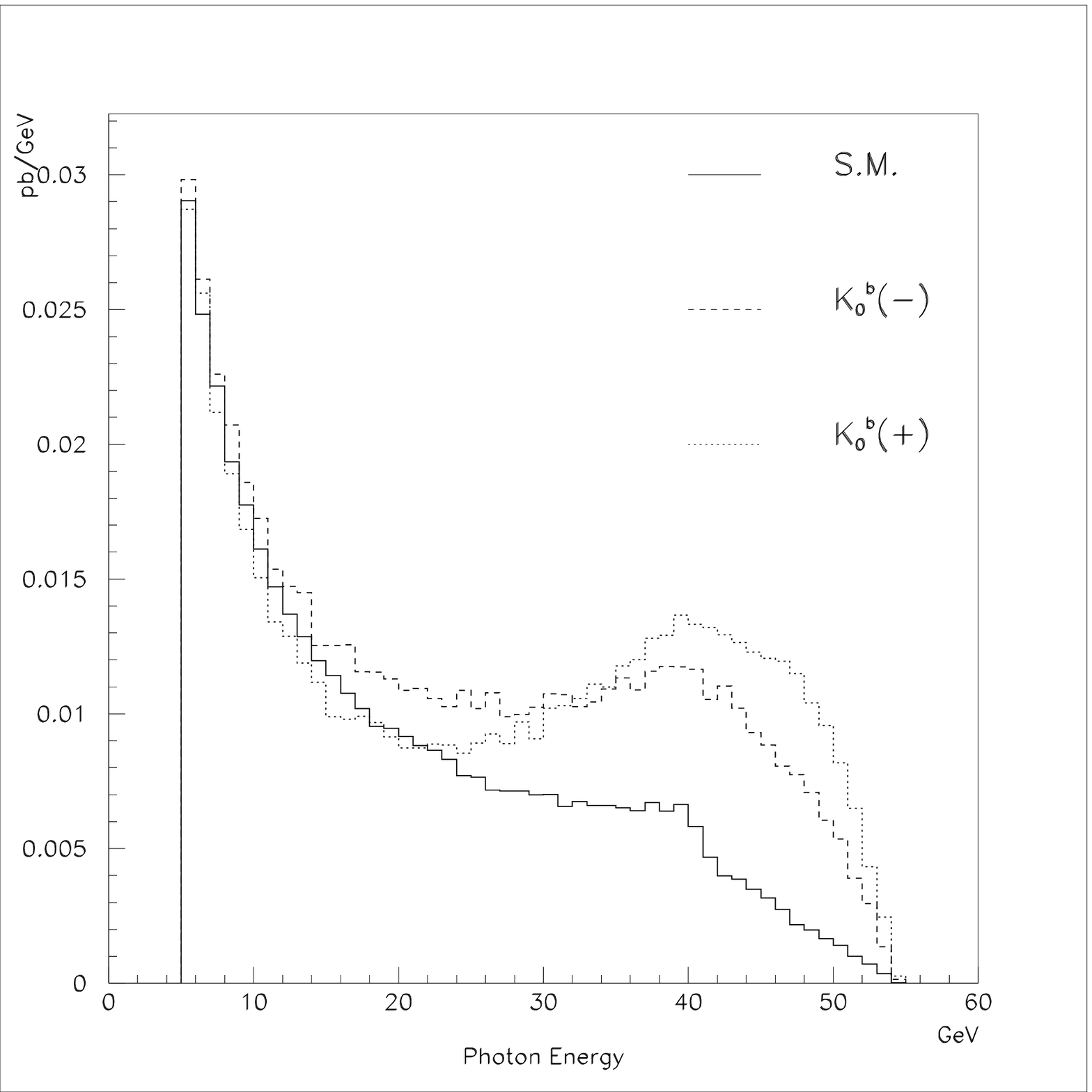}}
\mbox{\epsfxsize=8.5cm\epsfysize=10cm\epsffile{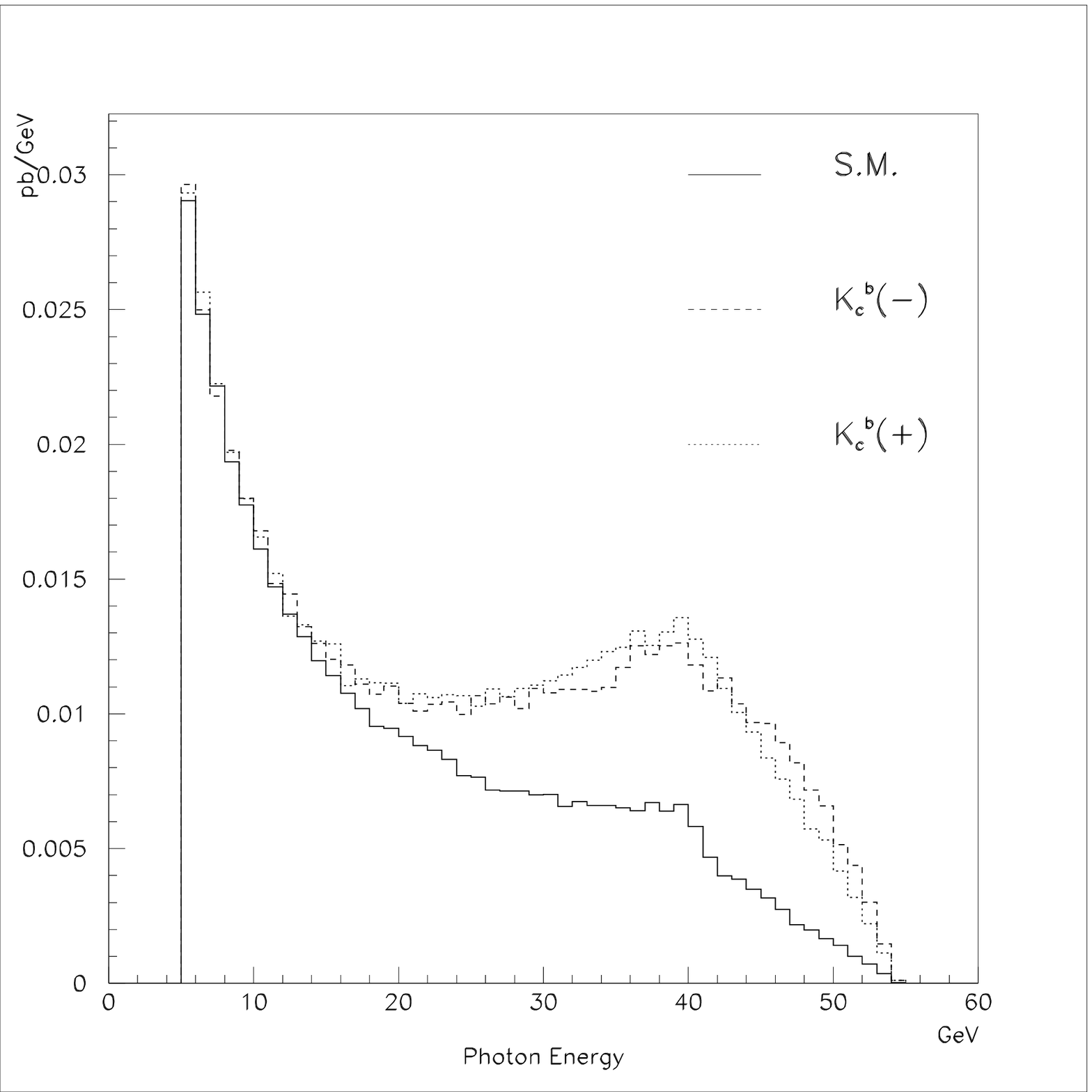}} }
\caption{\label{diszzg}{\em Distribution in the least energetic photon in
$\epem \ra \gamma \gamma Z$ due to the anomalous couplings
$k_0^b $ and $k_c^b$. We have taken $k_i$ values that lead to a
$3\sigma$ increase of the cross section as explained in the text.
The $\pm$ are the positive and negative values given by
Eq.~(\ref{limaazlep2}).
}}
\end{center}
\end{figure*}

Once again the most typical distribution is that of the least
energetic photon, as we can see from Fig.~\ref{diszzg}.
Here again, given enough statistics it should be possible to
disentangle between the two Lorentz structures.

\subsection{Improvement at high energy}
\begin{figure*}[htbp]
\begin{center}
\mbox{\epsfxsize=14cm\epsfysize=14cm\epsffile{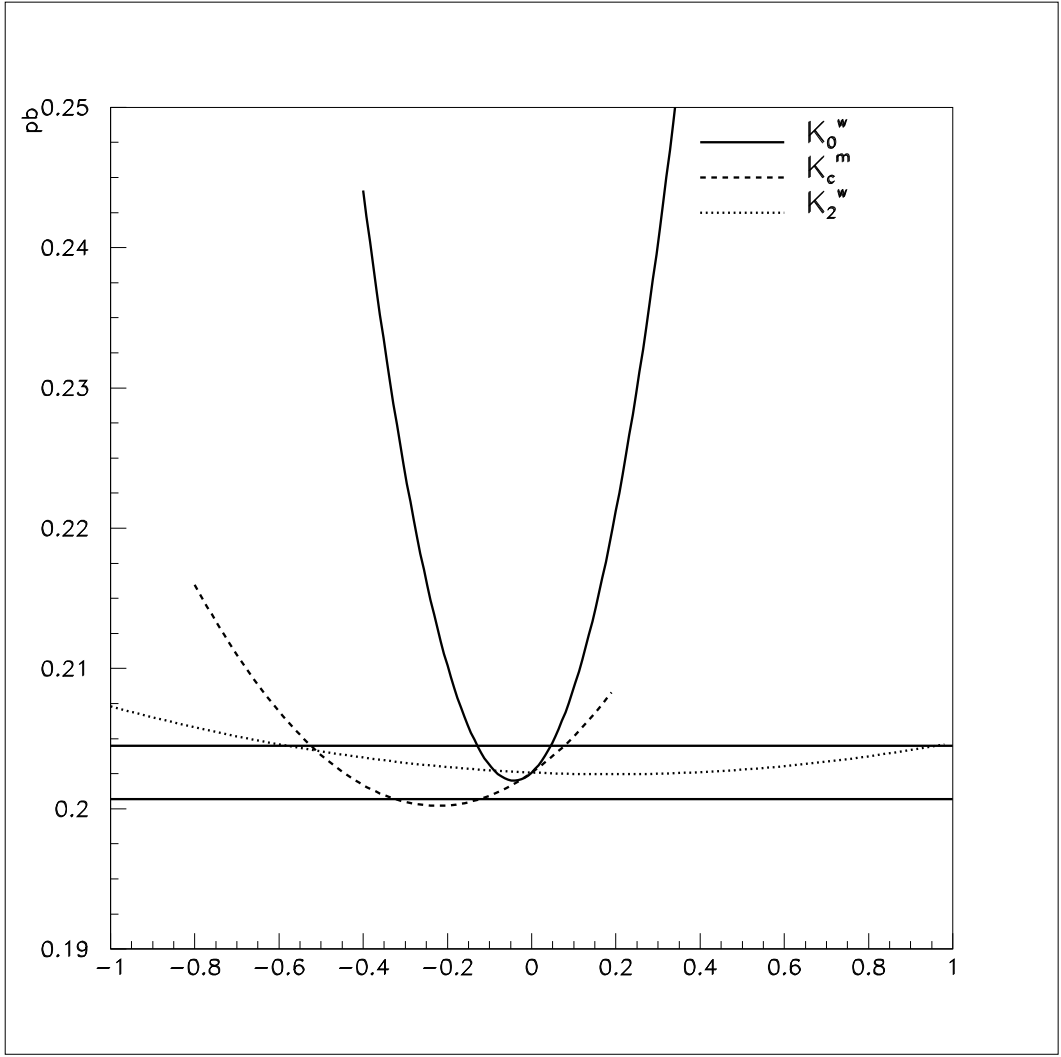}}
\caption{\label{ki-dependence-500}{\em As in Fig.~\ref{ki-dependence-lep2}
but with $\sqrt{s}=500$~GeV. Here both the $\pm 3\sigma$ deviation
with
${\cal{L}}=500$~fb$^{-1}$ are shown.
}}
\end{center}
\end{figure*}

All these couplings will be much better probed as the energy
increases. We have found that a linear collider  running at
$500$~GeV will improve these limits by as much as three orders of
magnitude, especially for the $k_0$ couplings, see
Fig.~\ref{ki-dependence-500}.  To extract the $3\sigma$ limits we
have assumed the same cuts as those at LEP2, we have only
concentrated on the use of the $WW\gamma$ channel where we find
the cross section to be $202.6$~fb. Choosing one operator from
each of the three sets, ($k_0,k_c,k_1$), and assuming a total
integrated luminosity of $500$~fb$^{-1}$, one will have the
following constraints

\beqn
-2.0\; 10^{-5}\; {\rm GeV}^{-2}&<&\frac{k_0^w}{\Lambda^2}<0.6\;
10^{-5} \;  {\rm GeV}^{-2}  \nonumber
\\ -8.1\; 10^{-5}\; {\rm
GeV}^{-2}&<&\frac{k_c^m}{\Lambda^2}<-5.0\; 10^{-5} \;  {\rm
GeV}^{-2} \;\; {\rm and} \;\; -2.0\; 10^{-5}\; {\rm
GeV}^{-2}<\frac{k_c^w}{\Lambda^2}<1.2\; 10^{-5} \;  {\rm GeV}^{-2}
\nonumber \\ -9.0\; 10^{-5}\; {\rm
GeV}^{-2}&<&\frac{k_2^w}{\Lambda^2}<15.0
\; 10^{-5} \;  {\rm
GeV}^{-2}
\eeqn

We have also analysed how the sensitivity on the above limits
changes if one increased the cut on the photon energy from $5$~GeV to
$20$~GeV. The limits hardly change.

Note that we can in principle also use other channels,
like for instance $\epem \ra W W Z$. However this channel is more
conducive to tests on the $WWZZ$ couplings which appear at a lower
order in the energy expansion in the context of the chiral
Lagrangian and are thus more likely.

\section{Remarks and conclusion}
We have given an extensive list of \cviol and \pviol conserving
quartic bosonic operators involving a photon and which may be
probed at LEP2. Previous studies have considered only two
operators beside a third which we have shown to be \cpviol
violating. We have shown how these structures can be embedded in a
fully $SU(2) \times U(1)$ and $SU(2)_c$ globally invariant
operators. We have derived limits on these couplings from \eewwgt
and $\epem \ra Z\gamma\gamma$ at LEP2. When constraining the
different structures so that we reproduce the contrived models
considered by the OPAL collaboration \cite{Opalquartic1} and in
\cite{Stirlingquartic}, our limits are consistent with theirs. We
should however add a note of warning. The natural size of these
couplings, $k_i^j$, should be of order unity for $\Lambda \sim 4
\pi v \sim 3$TeV. Viewed this way the limits one will extract from
LEP2 are not very  meaningful and are much worse compared to the
limits on the tri-linear couplings derived from LEP2.  However the
next generation of linear colliders can quite usefully constraint
these operators, since we can gain as much as three-orders of
magnitude compared to LEP2. Some order of magnitude on these
non-renormalisable operators can also be set from their
contributions to the low-energy precision measurements. However a
study within a fully gauge invariant framework has not been done.
A partial investigation \cite{eboliquartic} taking into account
only two operators, with the restriction that no $WWZ\gamma$ and
$ZZZ\gamma$ ensue, has been attempted, however the approach taken
in \cite{eboliquartic} leads to these operators not decoupling in
loop contributions and therefore cast a shadow on limits derived
this way. For a discussion of how to treat the loop effects of the
anomalous operator on low energy observables one should refer to
\cite{Cliff,Szalapski}.

\setcounter{section}{0}

\setcounter{subsection}{0}
\setcounter{equation}{0}
\def\thesection{\Alph{section}}
\def\thesubsection {\thesection.\arabic{subsection}}
\def\theequation{\thesection.\arabic{equation}}

\setcounter{equation}{0}
\def\thequation{\thesection.\arabic{equation}}

\section{Appendix}
In \cite{Stirlingquartic} a $WWZ\gamma$ operator not listed in
Eq.(~\ref{coup_wwzg})  is also considered. Though $SU(2)_c$
symmetric, it is explicitly \cpviol violating. The authors
\cite{Stirlingquartic} take

\beqn
\label{ebolinoiw}
{{\cal L}}_{n}= -\frac{e^2}{16 \Lambda^2} a_n \eps_{ijk}
W^{i}_{\mu \alpha} W^{j}_\nu W^{k\; \alpha}
 F^{\mu \nu}
\eeqn

where the $W^{i}$ are the elements of $W$ triplet before mixing.
Note that this Lagrangian differs from that of \cite{eboliquartic}
by an overall $i$ factor which would make it non hermitian.
Expanding in the physical fields one would get:

\beqn
\label{ebolinoiwz}
{{\cal L}}_{n}  & =&\;- i\; \frac{e^2}{16 c_W \Lambda^2} a_n
\left\{ F_{\mu}^{\;\;\;\nu} \left[ Z^{\mu \alpha} (W^+_\alpha W^-_\nu
- W^-_\alpha W^+_\nu) \right. \right. \nonumber
\\  &+& \left. W^{+\;\mu \alpha} (W^-_\nu
Z_\alpha -Z_\nu W^-_\alpha) \;-\; W^{-\;\mu \alpha} (W^+_\nu
Z_\alpha -Z_\nu W^+_\alpha) \right]
\eeqn

Note now that properly going to the physical basis, the Lagrangian expressed in terms of
the charged fields has an $i$ as required by hermiticity. In the passage from
Eq.~\ref{ebolinoiw} to Eq.~\ref{ebolinoiwz}, a $i$ is missing in \cite{Stirlingquartic}.
As explicit in Eq.~\ref{ebolinoiwz} this $i$ is crucial for hermiticity. On the other
hand it is quite explicit also that this coupling violates \cviol and \cpviol. Even if one had
considered this coupling in computing $\epem \ra W^+ W^- \gamma$, without any (transversely)
polarized
beams or the study of specific correlations in the decay products,
this couplings does not interfere with the \sm amplitudes.
Therefore one only has a quadratic sensitivity on this anomalous coupling It is also interesting
to write this Lagrangian in a gauge invariant manner. For instance in the chiral
Lagrangian approach we may write:

\beqn
{{\cal L}}_{n}=i \frac{a_4^{{\rm CP}}}{\Lambda^2} g'\;g \tr
\left(\tau_3 \B_{\mu \nu}\right) \; \tr \left(\W^{\mu \alpha}
\left[ \V_\nu, \V_\alpha \right] \right)
\eeqn

When written in terms of the physical fields this leads to the
quartic couplings
\beqn
{{\cal L}}_{n} \rightarrow & &\frac{i}{2}\; e \; g^2 \; g_Z \;
\frac{a_4^{{\rm CP}}}{\Lambda^2} \left\{ F_{\mu}^{\;\;\nu} \left[
Z^{\mu \alpha} (W^+_\alpha W^-_\nu - W^-_\alpha W^+_\nu) \right.
\right. \nonumber
\\  &+& \left. W^{+\;\mu \alpha} (W^-_\nu
Z_\alpha -Z_\nu W^-_\alpha) \;-\; W^{-\;\mu \alpha} (W^+_\nu
Z_\alpha -Z_\nu W^+_\alpha) \right] \nonumber
\\  &-& \left.  \frac{s_W}{c_W} Z_{\mu}^{\;\;\nu}\left[ W^{+\;\mu \alpha} (W^-_\nu
Z_\alpha -Z_\nu W^-_\alpha) \;-\; W^{-\;\mu \alpha} (W^+_\nu
Z_\alpha -Z_\nu W^+_\alpha) \right] \right\}
\eeqn

we would then have with the correct $i$,
\beqn
a_n=\;-\;\frac{8}{s_W} g^2 a_4^{{\rm CP}}
\eeqn

Note that this vertex when written in a gauge invariant manner
also contributes to a $WWZZ$ vertex.


\end{document}